\documentclass[12pt]{article}

\usepackage[english]{babel}
\usepackage{amssymb,amsmath}
\usepackage{amsfonts}
\usepackage{latexsym}
\usepackage{verbatim}
\usepackage{amsthm}
\usepackage{amsbsy}
\usepackage{graphicx,color}
\usepackage{epsfig}
\usepackage{cite}
\usepackage[debug,pageanchor=false]{hyperref}

\definecolor{link}{rgb}{.8,.15,.1}
\hypersetup{colorlinks=true,linkcolor=link,citecolor=link,urlcolor=link,linktocpage}
\newcommand{\be}{\begin{equation}}
\newcommand{\ee}{\end{equation}}
\newcommand{\bi}{\begin{itemize}}
\newcommand{\ei}{\end{itemize}}
\newcommand{\bea}{\begin{eqnarray}}
\newcommand{\eea}{\end{eqnarray}}
\newcommand{\ba}{\begin{array}}
\newcommand{\ea}{\end{array}}

\def\Tr{\mathrm{Tr}}

\newcommand{\nn}{\nonumber}
\newlength{\sswidth}

\begin{document}
\begin{titlepage}

\vskip 1cm
\centerline{{\Large \textbf{Topological anomalies for Seifert 3-manifolds}}}

\vspace{1cm}

\vspace{1cm}

\centerline{    
  \textsc{  ~Camillo Imbimbo$^{1,2,a}$, ~Dario Rosa$^{3,4,5,b}$}  }

\vspace{0.5cm}

\begin{center}

$^1\,$ Dipartimento di Fisica, Universit\`a di Genova,
Via Dodecaneso 33, 16146 Genoa, Italy\\
 $^2\,$ INFN, Sezione di Genova, Via Dodecaneso 33, 16146, Genova, Italy\\
\vspace{0.3 cm}
$^3\,$ School of Physics and Astronomy \& Center for Theoretical Physics Seoul National University, Seoul 151-747, Korea\\
$^4\,$ Dipartimento di Fisica, Universit\`a di Milano-Bicocca, I-20126 Milano, Italy\\
$^5\,$ INFN, Sezione di Milano-Bicocca, I-20126 Milano, Italy
		    
\end{center}

\begin{center}
{\small $^a$camillo.imbimbo@ge.infn.it, ~~$^b$dario.rosa85@snu.ac.kr}
\end{center}

\vspace{1cm}

\centerline{\textsc{ Abstract}}
\vspace{0.2cm}
 {\small
 We study globally supersymmetric 3d gauge theories on curved manifolds by 
describing the coupling of 3d topological gauge theories, with both Yang-Mills and Chern-Simons  terms in the action,  to background topological gravity.  In our approach, the Seifert condition for manifolds  supporting global  supersymmetry is elegantly deduced from the BRST transformations of topological gravity. A cohomological characterization of the geometrical moduli which affect the partition function is obtained.  In the Seifert context the Chern-Simons topological (framing) anomaly is BRST trivial. We compute explicitly the corresponding local Wess-Zumino functional. As an application,   we
obtain the dependence  on the Seifert moduli of the partition function of 3d supersymmetric gauge theory on the squashed sphere by solving the anomalous topological Ward identities, in a regularization independent way and without the need of evaluating any functional determinant.
}

\thispagestyle{empty}

\vfill
\eject
\end{titlepage}
\hypersetup{pageanchor=true}

\tableofcontents
\section{Introduction} 

\label{sec:intro}

In the last few years there has been considerable progress in the analytical evaluation of partition functions
and observables of supersymmetric gauge theories in different dimensions on certain  compact manifolds
equipped with suitable metrics. The common theme of these computations is localization.  Localization
is  a long-known property of  supersymmetric and topological theories,  by virtue of which semi-classical approximation becomes, in certain cases, exact \cite{Witten:1991zz}.  In more recent times this property has been exploited with considerable success in the work by  Pestun \cite{Pestun:2007rz}  and  in many following papers. In Pestun's approach no twisting of supersymmetry is performed.  One rather seeks for manifolds and metrics 
 supporting (generalized) covariantly constant spinors which ensure that certain supersymmetry global charges are unbroken. The global supersymmetry
charges, even if spinorial in character, function essentially as  topological BRST charges.   Under favourable conditions
one can choose a Lagrangian for which the semi-classical computation in the supersymmetric background  is exact.

In three dimensions, a host of results is available. Explicit exact computations have been performed for 3-spheres, both with round and ``squashed'' metrics, and for Lens spaces. The best understood case is the one  for which the complex conjugate of the (generalized) covariantly constant spinor is also covariantly constant. This is referred to as the ``real'' case in \cite{Closset:2012ru}. In all these cases  the existence of (generalized) convariantly constant spinors implies in turn the existence of a {\it Seifert structure} on the 3-manifold. 
This refers to 3-manifolds with an almost contact metric structure and associated Reeb  Killing vector field. 

As a matter of fact
Seifert 3-manifolds had already made their appearance earlier,  in the study of non-supersymmetric pure Chern-Simons (CS) gauge theories.
It was discovered  first ``experimentally'' \cite{Rozansky:1993zx}  and then explained using various approaches by different authors  \cite{Beasley:2005vf},\cite{Blau:2006gh} that the semiclassical approximation for CS theories becomes exact precisely for Seifert 3-manifolds.  Later, starting from  \cite{Kapustin:2009kz}, this result was rederived by considering  the supersymmetric extension of CS: indeed this model
 is equivalent, after integrating out non-dynamical auxiliary fields, to the bosonic theory.  In this way, computability of CS on Seifert manifolds was brought within the more general paradigm of convariantly constant spinors and localization. 

In some cases, it is possible to perform localization computations not just for a single isolated Seifert structure,
but for families of Seifert metrics depending on some continuous parameters.
A significant example is provided by the squashed metric on the 3-sphere \cite{Hama:2011ea}.\footnote{ In this paper,  we will consider the squashing of the 3-sphere that in \cite{Hama:2011ea} is referred to as the ``less familiar'' one. This is the squashing that preserves only  the $U(1)\times U(1)$ isometry.
A related deformation of the round sphere,   the branched sphere $S^3_q$, is  discussed in \cite{Huang:2014gca}.
}
In those instances  the partition function (and the observables) turns out to depend non-trivially
on (some of) those parameters. Take for example the case of (supersymmetric) Chern-Simons theory  on the squashed spheres, with the metric  
\bea
ds^2 = \bar{g}_{\mu\nu}(x; b)\, dx^\mu\otimes dx^\nu=(\sin^2\theta+ b^4\,\cos^2\theta)\, d\theta^2 +\cos^2\theta\, d\phi_1^2 + b^4\,\sin^2\theta\, d\phi_2^2 \ .
\label{squashedmetricintro}
\eea
At first sight, the fact that the partition function is  a non-trivial function of the squashing parameter $b^2$ is not, {\it per se}, surprising. Indeed, even if CS theory is topological at classical level, topological
invariance is anomalous at quantum level \cite{Witten:1988hf}. This means that the quantum CS action does depend on the background
metric in a way which is controlled by the anomaly functional:
\bea
\delta \Gamma_{CS}[ g_{\mu\nu}]  =  \frac{c}{6} \int_{M_3} A^{(3)}_1[g_{\mu\nu}, \psi_{\mu\nu}]=\frac{c}{6}\, \int_{M_3} \epsilon^{\mu\nu\rho}\, R^\alpha_\mu
\, D_\nu\,\psi_{\rho\alpha}\, d^3x \ ,
\label{topanomalyintro}
\eea
where $\psi_{\mu\nu}=\delta g_{\mu\nu}$ is the variation of the metric and $c$ is a computable anomaly coefficient.
However, if one plugs both the squashed metric (\ref{squashedmetricintro}) and its variation $\psi_{\mu\nu}=b\,\partial_b\, g_{\mu\nu}$ into the anomaly form (\ref{topanomalyintro}), one finds that the
anomaly vanishes identically
\bea
 A^{(3)}_1[g_{\mu\nu}(x; b), b\,\partial_b\, g_{\mu\nu}]=0 \ .
 \label{topanomalysquashedintro}
\eea
In this work we will solve this conundrum: we will see that  the vanishing of the topological anomaly 
for the squashed spheres is compatible with the non-trivial dependence of the partition function
on the squashing parameter. As a matter of fact, we will show that  the topological anomaly   captures the precise dependence of the partition function on $b$.  We will extend this results to
generic three-dimensional supersymmetric theories, with both Yang-Mills and Chern-Simons terms in the action (YM+CS),  involving vector multiplets only.

The resolution of our puzzle will require understanding the appropriate renormalization prescription  for  quantum effective actions on Seifert manifolds.
The time-honored method to identify the renormalization prescriptions associated to certain symmetries is to introduce backgrounds fields which act as sources for the currents associated to those symmetries. 
This approach has been forcefully advocated more recently in the specific context of supersymmetric gauge theories in \cite{Festuccia:2011ws}  and in several following papers.

We also  will introduce backgrounds, but our treatment will differ from the one which has become common in
the literature on localization of the last few years. Instead  of coupling the supersymmetric gauge theory to supergravity, we will  first consider its topological version and  then couple it to {\it topological gravity}. 

This will have  the advantage of obtaining the Seifert condition for global supersymmetry in the most straightforward way by avoiding all the complications of spinors. Moreover and most importantly the topological gravity
formulation will make immediate to identify the subsets of the topological transformations which preserve
the Seifert structure. These topological Seifert BRST transformations lead to a cohomological characterization of the geometrical moduli which affect the partition function. We compute the cohomology which parametrizes Seifert moduli in Section \ref{Section:moduli}. 

 In the Seifert context the topological (also known as ``framing'') anomaly is BRST trivial \cite{Beasley:2005vf}. In this paper we compute explicitly the corresponding local Wess-Zumino (WZ) functional. 
To our knowledge, this has not been done before.
 We will call this functional the {\it Seifert WZ action} and write it down in Eq. (\ref{WZSeifertaction}).  The Seifert WZ action is a local functional which is invariant under Seifert-preserving reparametrizations  and whose BRST variation equals the topological anomaly for Seifert manifolds.  We will use the Seifert WZ  action to  derive the dependence on the Seifert moduli  of the quantum action  directly from  the anomalous Ward identity associated to  the topological anomaly. Our computation will be manifestly regularization and gauge independent: We will do it without the need of  computing explicitly any functional determinant.

Let us clarify the relationship between our Seifert WZ action and the CS gravitational functional.  It is well known that the topological anomaly (\ref{topanomalyintro}) can be expressed, for {\it generic} (i.e. also non-Seifert)  3-manifolds,  as a BRST variation 
\bea
  A^{(3)}_1 = s\, \Gamma_{GCS}^{(3)}[g]
  \label{GCSintro}
\eea
where $\Gamma_{GCS}^{(3)}[g]$ is the 3-dimensional gravitational Chern-Simons action
\bea
 \Gamma_{GCS}^{(3)}[g] = \mathrm{Tr}\, \bigl( \frac{1}{2}   \Gamma\, d\,\Gamma + \frac{1}{3}\, \Gamma^3\bigr)
 \label{GCSaction}
\eea
$ \Gamma_{GCS}^{(3)}[g]$ is not a globally defined 3-form  and thus  it is not a legitimate Wess-Zumino local action.  This is why  $ A^{(3)}_1$ is a genuine anomaly (for generic 3-manifolds).  It is possibile however to trade full\footnote{By full  reparametrization invariance we mean invariance under both ``small'' and ``large'' coordinate transformations.} reparametrization invariance with topological invariance by considering the renormalization prescription in which one subtracts the gravitational CS action from the non-local quantum action $\Gamma_{CS}[ g]$.  The resulting renormalized action
\bea
\tilde{\Gamma}_{CS}[g]= \Gamma_{CS}[ g]- \Gamma_{GCS}^{(3)}[g]
\label{Gammatopologicalintro}
\eea
is topological (i.e. it is independent of  the metric) but it is not fully reparametrization invariant: it depends on the framing.  The fact that the topological
anomaly, when evaluated for the squashed sphere background, vanishes (Eq. (\ref{topanomalysquashedintro})),
means that the gravitational Chern-Simons action, when evaluated on the same background, is a constant
independent of $b$.  Hence Eq.  (\ref{Gammatopologicalintro}) implies that  also the non-local quantum action $ \Gamma_{CS}[ g]$ evaluated for  the squashed sphere  is $b$-independent.  In other words the gravitational CS action does {\it not} capture the dependence of
the partition function on the squashing parameter.  As we will explain in detail in Section \ref{Section:seiferttopanomaly}, the reason for this is that  the gravitational CS action is not the counterterm appropriate for renormalizing supersymmetric  theories on Seifert manifolds. The correct counterterm is the one which respects the relevant --- reparametrization and topological --- Seifert simmetries: and it is precisely our Seifert Wess-Zumino action.  In Appendix \ref{App:AppendixCGSWZ} we show
that the gravitational CS action and the Seifert Wess-Zumino action differ by a total derivative  which does not integrate to zero on Seifert manifolds.

In the supergravity context, the gravitational CS action (\ref{GCSaction}) admits a supersymmetric extension which has been discussed in \cite{Closset:2012vg}.  We will show in Appendix \ref{App:AppendixSUGRA} that also this supergravity CS local action does not capture the correct
dependence on the Seifert moduli, for the same reason mentioned in the paragraph above: the counterterm which respects  the Seifert symmetry is a supersymmetrization of our Seifert WZ  action (\ref{WZSeifertaction})  which involves both gravitational and  gauge fields. The correct $b$-dependence of the quantum action is obtained by taking the
difference of the supergravity CS action evaluated for the squashed background and the supersymmetric Seifert WZ  action.  In the same appendix we also show that the gauge counterterm which supersymmetrize the Seifert WZ  action  (\ref{WZSeifertaction}) does not contribute to the overall moduli dependence of the partition function. The full moduli dependence is still captured by the same
Seifert WZ action which emerges in the topological gravity framework.  In this way we verify directly that both the topological and the supergravity language lead to the same answer.

\section{Coupling 3d Chern-Simons to topological gravity} 
\label{sec:3dCS}

The classical Chern-Simons action \cite{Schwarz:1978cn} is
\be
\Gamma_{CS} =  \int_{M_3} {\rm Tr}\Big[ \frac{1}{ 2}\,A\,d\,A +\frac{1}{ 3} A^3\Bigr] \ ,
\label{classicalaction}
\ee
where
\be
A = A^a_\mu\, T^a\, dx^\mu 
\ee
is a 1-form gauge field on a closed 3-manifold $M_3$. $T^a$, with
 $a=1,\ldots, {\rm dim}\, G$, are generators of
the Lie algebra of the simple, connected and simply connected gauge group $G$. Gauge invariance leads to the nilpotent BRST transformation rules\footnote{We will adopt the convention that BRST operator $S_0$ {\it anti-commutes} with
the exterior differential $d$.}
\bea
&& S_0 \, A = - D\, c \ ,\qquad S_0\, c = - c^2 \ ,
\label{brszerofields}
\eea
where $c = c^a\, T^a$ is the ghost field carrying ghost number $+1$ and
$D\, c\equiv d\,c +[A, c]_+$ is the covariant differential.

The classical action (\ref{classicalaction}) is both invariant under diffeomorphisms and independent of the 3-dimensional background metric $g_{\mu\nu}$. In order to study the fate in the quantum theory of this {\it global} topological symmetry 
one must couple  the theory to suitable backgrounds.  This has been done in \cite{Imbimbo:2009dy} where it is explained that the backgrounds appropriate for the
topological symmetry in question are those of equivariant
{\it topological gravity} \cite{Baulieu:1988xs}
\bea
&& s \,g_{\mu\nu} = \psi_{\mu\nu} -{\cal L}_\xi\, g_{\mu\nu} \ ,\qquad
s \,\psi_{\mu\nu} = {\cal L}_\gamma\, g_{\mu\nu}-{\cal L}_\xi\, 
\psi_{\mu\nu} \ ,\nonumber\\
&&s \,\xi^\mu = \gamma^\mu -\frac{1}{2}\, {\cal L}_\xi\, \xi^\mu \ ,\qquad
 s \,\gamma^{\mu} =-{\cal L}_\xi\,\gamma^\mu \ ,
\label{brsgravityequivariantintro}
\eea
where $\xi^\mu$ is the ghost of reparametrizations of ghost number +1, $\psi_{\mu\nu}$ is the topological gravitino of ghost number +1 and $\gamma^\mu$ is the ghost-for-ghost of ghost number +2. 

The coupling to background topological gravity induces both deformations in the BRST transformations of the matter fields and extra terms in the action. It also requires introducing new matter fields $\tilde{A}$ and $\tilde{c}$ which sit in the same BRST multiplet as $c$ and $A$. $\tilde{A}$ and $\tilde{c}$  are  
Lie algebra-valued 2 and 3-forms
\bea
&&\tilde{A} = (\tilde{A})^a_{\mu\nu}\, T^a\, dx^\mu\,dx^\nu \ ,\qquad 
\tilde{c} = (\tilde{c})^a_{\mu\nu\rho}\, T^a\, dx^\mu\,dx^\nu\, dx^\rho \ ,
\eea
of ghost number $-1$ and $-2$ respectively.  All the matter fields fit niceley into a single super-field, or polyform,
$\mathcal{A}$, 
\bea
\mathcal{A} \equiv c + A + \tilde{A} + \tilde{c}
\label{CSsuperform}
\eea 
whose total fermionic number, given by the form degree plus ghost number, is +1.\footnote{The benefit of  polyforms for analyzing the BRST structure of flat Chern-Simons theory was first pointed out in  \cite{Axelrod:1993wr}.}  
The action of the BRST transformation on the supermultiplet (\ref{CSsuperform}) {\it before} coupling to topological
gravity writes in the compact form
\bea
\delta_0 \,\mathcal{A} + \mathcal{A}^2=0 \ ,
\label{CSrigid}
\eea
where
\bea
\delta_0 = S_0+ d \ ,\qquad \delta_0^2=0
\label{rigiddelta}
\eea
is the ``rigid''  coboundary operator of total fermion number +1.  It is immediate to check that (\ref{CSrigid}) reduces
to (\ref{brsgravityequivariantintro}) when restricted to the fields $c$ and $A$.  

To describe the coupling to topological gravity it is convenient to consider the operator
 \bea
 S \equiv s + \mathcal{L}_\xi \ ,
 \eea
rather than the nilpotent BRST operator $s$. On the functionals of the backgrounds independent of $\xi^\mu$, $S$ satisfies
 \bea
 S^2 = \mathcal{L}_\gamma
 \eea
and it is  therefore nilpotent on the space of equivariant functionals of the topological gravity multiplet. 

After these preliminaries, one discovers that  the coboundary operator appropriate to describe the BRST transformation rules of the system {\it after} coupling  to topological gravity is simply 
\bea
\delta \equiv S + d- i_\gamma \ ,\qquad \delta^2=0 \ ,
\label{tgdelta}
\eea
where $\delta$  writes
in the same form as the rigid ones
\bea
\delta\,\mathcal{A} + \mathcal{A}^2=0 \ .
\label{brsmatterequivariantfixedbis}
\eea
When written for the component fields, these  transformations become
\bea
&&s\, c = -c^2 - {\cal L}_\xi\, c +i_\gamma(A) \ ,\nonumber\\
&&s\, A = -D\, c  - {\cal L}_\xi\, A + i_\gamma (\tilde{A}) \ ,\nonumber\\
&&s\, \tilde{A} =- [\tilde{A},c]- {\cal L}_\xi\, \tilde{A}-  F+i_\gamma(\tilde{c}) \ ,\nonumber\\
&&s\,\tilde{c} =-[\tilde{c},c] - {\cal L}_\xi\, \tilde{c} - D\,\tilde{A} \ .
\label{brsmatterequivariant}
\eea
The action also gets modified after coupling to topological gravity: one can check that 
\bea
&&\Gamma_{CS+t.g.}= \Gamma_{CS} + \frac{1}{ 2}\,\int_{M_3} \Tr\, i_\gamma(\tilde{A})\, \tilde{A} \ ,
\label{actionCSbis}
\eea
 is invariant  by transforming both the fields according to   (\ref{brsmatterequivariant}) and the backgrounds according to   (\ref{brsgravityequivariantintro}).

The fields  $\tilde{A}$ and $\tilde{c}$ have a natural
interpretation, in the Batalin-Vilkovisky formalism,  as the anti-fields of $A$ and $c$.  In this traditional BV framework, which is the one adopted in   \cite{Axelrod:1993wr},\cite{Imbimbo:2009dy},  the action $\Gamma$ should be supplemented with the canonical piece
for all the fields and backgrounds:
\bea
&& \Gamma_{\mathrm{can}}= \int_{\Sigma} s\,c\, \tilde{c} + s\, A\, \tilde{A} + s\, \tilde{A}\, \tilde{A} + s\, \tilde{c}\, \tilde{c}+ s\,g_{\mu\nu}\, \tilde{g}^{\mu\nu} + s\,\psi_{\mu\nu}\, \tilde{\psi}^{\mu\nu} + s\,\gamma^\mu\, \tilde{\gamma}_\mu \ .
\label{canonicalActionCSBV}
\eea
The full BV action 
\bea
&&\Gamma_{BV} = \Gamma_{CS+t.g.} + \Gamma_{\mathrm{can}} \ ,
\eea
generates the BRST transformations of both fields and anti-fields via the familiar BV formulas.

However in this paper  we will argue that an alternative --- although exotic ---  point of view is available in our context. In the approach that we propose  $\tilde{A}$ and $\tilde{c}$  are  {\it independent} auxiliary fields whose  function is to close the BRST transformations {\it off-shell}: at the same time, our action we will be just $\Gamma_{CS+t.g.}$,  not the full $\Gamma_{BV} $.  

This approach is consistent since the BRST transformations close on the fields $(c, A, \tilde{A},\tilde{c})$ and leave  $\Gamma_{CS+t.g.}$ invariant. Of course, in the formulation in which $(\tilde{A},\tilde{c})$ are auxiliary fields, $\Gamma_{CS+t.g.}$ maintains the original non-abelian gauge symmetry, which, eventually, will have to be fixed: both the gauge non-abelian symmetry and the global vector supersymmetry  associated to the topological invariance of the matter theory are captured by the BRST symmetry in (\ref{brsmatterequivariant}).  In other words, the anti-fields of the BV formulation can be reinterpreted as the auxiliary fields  which are necessary to close the global supersymmetry algebra of the supersymmetric CS, {\it after} twisting that supersymmetry to obtain a topological model. 

As a matter of fact, the 3d action (\ref{actionCSbis}) has {\it more} local gauge invariance than just  the standard non-abelian gauge invariance: it  is invariant also under the fermionic local symmetry
\bea
\tilde{A} \to \tilde{A} + i_\gamma(\chi) \ ,
\eea
where $\chi$ is a fermionic scalar  gauge parameter in the adjoint of the gauge group. 
Thus, the  commuting field $\tilde{c}$ can be viewed as  the ghost associated to this additional local symmetry.  This extra gauge invariance is  fixed by replacing  $\Gamma_{CS+t.g.}$ with
\bea
&&\Gamma^\prime_{CS+t.g.} = \Gamma_{CS+t.g.} + s\,\int_{M_3} \Tr \bigl( b\ast i_\gamma( \ast \tilde{A})\bigr)= \nn\\
&&\qquad =\Gamma_{CS} + \frac{1}{ 2}\,\int_{M_3} \Tr\, i_\gamma(\tilde{A})\, \tilde{A}+\int_{M_3} \Tr \bigl[ \Lambda\ast i_\gamma( \ast \tilde{A})- b\ast i_\gamma( \ast \,i_\gamma (\tilde{c}))+ b\ast i_\gamma( \ast \,F)\bigr]+\nn\\
&&\qquad\qquad+ \int_{M_3} d^3 x\,\psi_{\mu\nu}\, \epsilon^{\alpha\beta\mu}\,\gamma^\nu\, \Tr\, b\, \tilde{A}_{\alpha\beta}\,  \ ,
\eea
where $b$ is a 0-form anti-ghost of ghost number -2,
\bea
s\, b =- \mathcal{L}_\xi\,b -[c, b]+ \Lambda \ ,\qquad s\, \Lambda =-\mathcal{L}_\xi \, \Lambda- [c, \Lambda]+ i_\gamma(D\, b) \ ,
\eea
$\Lambda$ is a Lagrange multiplier of ghost number -1 and  $\ast$ is the Hodge dual with respect to the background metric $g_{\mu\nu}$.

Summarizing: The action $\Gamma^\prime_{CS+t.g.}$ has the background  topological supersymmetry captured by  (\ref{brsgravityequivariantintro}) and (\ref{brsmatterequivariant}) 
and no other local gauge-invariance beyond the standard non-abelian gauge invariance: $\tilde{A}$ and $\tilde{c}$ can now be consistently thought of as  auxiliary, non propagating,  fields  rather than anti-fields. 
\section{Coupling 3d topological YM  with CS term to topological gravity} 
\label{sec:3dYM}

The 3d topological YM theory is characterized by the BRST transformations
\bea
&& S_0 \, c= -c^2 + \sigma \ ,\nn\\
&&S_0\, A = -D\,c + \Psi \ , \nn\\
&& S_0\, \Psi = -[c, \Psi] - D\,\sigma \ , \nn\\
&& S_0\, \sigma = -[c,\sigma] \ ,
\label{brsYM}
\eea
$\Psi$ is a fermionic 1-form of ghost number 1  and $\sigma$ a bosonic 0-form
of ghost number 2. Both of them are in the adjoint of the gauge group.  It is convenient to introduce a super-field
or polyform of total fermionic number (ghost number + form degree) equal to 2:
\bea 
\mathcal{F} = F + \Psi + \sigma \ .
\label{YMsuperform}
\eea
The transformations (\ref{brsYM}) write in a nice compact form in terms of the ``rigid'' coboundary operator
$\delta_0 = S_0 +d$
\bea
\mathcal{F} =  \delta_0 \, \mathcal{A}_{YM}+ \mathcal{A}^2_{YM} \ ,
\eea
where
\bea
\mathcal{A}_{YM} = c + A \ .
\label{YMpoly}
\eea
It is important to observe that the super-field or polyform containing the gauge connection which is appropriate for the YM theory is not the same as the connection polyform $\mathcal{A}$ (\ref{CSsuperform}) of CS.

Let us again denote by $s$ the nilpotent BRST operator after coupling to topological gravity. As seen in the previous section, it is convenient to introduce the operator
\bea
 S_{YM} \equiv s + \mathcal{L}_\xi \ , \qquad S_{YM}^2 = \mathcal{L}_\xi \ ,
 \eea
 where $\xi$ is the ghost associated to reparametrizations.  The coboundary operator for topological YM coupled
 to topological gravity is again given by a formula identical to (\ref{tgdelta})
 \bea
\delta_{YM}\equiv S_{YM} + d- i_\gamma \ ,\qquad \delta_{YM}^2=0 \ ,
\eea
 with $\delta_{YM}$ satisfying 
 \bea
\mathcal{F} =  \delta_{YM} \, \mathcal{A}_{YM} + \mathcal{A}^2_{YM} \ .
\label{brsYMtg}
\eea
 These transformations write in components:
 \bea
 &&S_{YM}\, c = -c^2 + i_\gamma( A) +\sigma \ , \nn\\
&&S_{YM}\, A = -D\,c + \Psi \ , \nn\\
&& S_{YM}\, \Psi = -[c, \Psi] + i_{\gamma} (F) - D\,\sigma \ , \nn\\
&& S_{YM}\, \sigma = -[c,\sigma]+ i_{\gamma} (\Psi) \ .
\label{brsYMtgcomponents}
\eea
Finally, the action of pure topological YM consists only of a $s$-trivial term:
\bea
\Gamma_{YM+t.g.} = S_{YM}\, \chi \ .
\label{YMaction}
\eea
As just remarked, the superfield (\ref{YMpoly}) appropriate for YM theory is quite different from the corresponding polyform relevant for Chern-Simons theory.  However, we will now show that it is possibile
to recast the topological YM BRST transformations  purely in terms of the Chern-Simons superfield $\mathcal{A}$.
To this end, let us pick a contact structure $k$ on the 3-manifold, $M_3$ which is dual to the background vector field 
$\gamma^\mu$:
\bea
i_{\gamma} (k)=1 \ ,\qquad \mathcal{L}_ {\gamma}\, k=0 \ .
\eea
1-forms $\omega$ on $M_3$ are naturally decomposed along the horizontal and vertical directions as follows
\bea
\omega = k\, \omega_V + \omega_H \ ,
\eea
where
\bea
\omega_V\equiv i_{\gamma}(\omega) \ , \qquad i_{\gamma}(\omega_H)=0 \ .
\eea
Let us therefore decompose the fermionic 1-form $\Psi$ of the topological YM multiplet according to
\bea
\Psi \equiv  k\, \zeta +\Psi_H \ ,
\eea
with $i_{\gamma}(\Psi) =\zeta$.  Since any horizontal form is $i_{\gamma}$-exact
\bea
\Psi_H =  i_{\gamma}(\tilde{A}) \ ,
\eea
we have
\bea
\Psi \equiv  k\, \zeta +i_{\gamma}(\tilde{A}) \ ,
\label{contactdecomposition}
\eea
where $\tilde{A}$ is a 2-form of ghost number -1.  Note that the decomposition (\ref{contactdecomposition}) is not
unique and it has the gauge invariance
\bea
\tilde{A} \to \tilde{A} + i_\gamma(\tilde{c}) \ .
\label{ctildeinvariance}
\eea
When written in terms of $\zeta$ and $\tilde{A}$ the
YM topological transformations (\ref{brsYMtgcomponents}) rewrite as 
\bea
&&S_{YM}\, c = -c^2 + i_\gamma(A)+ \sigma \ ,\nn\\
&&S_{YM}\, A = -Dc +i_\gamma(\tilde{A})+ k\, \zeta \ ,\nn\\
&&S_{YM}\, \tilde{A} =  -[c, \tilde{A}] - F +i_\gamma(\tilde{c})+  k\, D\, \sigma \ , \nn\\
&&S_{YM}\,\tilde{c} = -[c,\tilde{c}] - D\,\tilde{A}- [\sigma, \tilde{A}]+ k\,dk\, \zeta \ ,\nn\\
&& S_{YM}\, \sigma = -[c,\sigma]+ \zeta \ ,\nn\\
&& S_{YM}\, \zeta = -[c,\zeta]+ i_{\gamma} (D\,\sigma) \ ,
\label{YMcontact}
\eea
where we introduced the ghost-for-ghost $\tilde{c}$ to take into account the gauge-invariance (\ref{ctildeinvariance}).
When written in this form, the emergence of the CS superfield
\bea
\mathcal{A} = \mathcal{A}_{YM} + \tilde{A}+ \tilde{c}
\eea
becomes apparent. Indeed the transformations (\ref{YMcontact}) can be expressed entirely in terms of $\mathcal{A}$:
\bea
\delta_{YM}\,\mathcal{A} + \mathcal{A}^2= \mathbf{\Phi} \ ,
\label{YMCScontact}
\eea
where  $\mathbf{\Phi}$ is the polyform of total ghost number +2
\bea
\mathbf{\Phi}= \sigma+ k\, \zeta + k\, D\,\sigma+ k\, \bigl(dk\, \zeta- [\sigma, \tilde{A}]) \ .
\eea
Eq. (\ref{YMCScontact}) is perfectly equivalent to the original (\ref{brsYMtgcomponents}): by means of  the decomposition
associated to the contact form $k$, it was possible to reformulate the topological YM transformations in 3d
 in terms of  the full CS multiplet.

The important observation now is that it is possible to recast (\ref{YMCScontact}) in the  CS-form
\bea
\delta_{YM} \,\mathcal{A}^\prime(\mathcal{A})+\mathcal{A}^\prime(\mathcal{A})^2=0 \ ,
\label{brstCSYM}
\eea
where
\bea
\mathcal{A}^\prime(\mathcal{A}) &\equiv& \mathcal{A} +\mathbf{\Theta} \ ,\nn\\
\mathbf{\Theta}& \equiv & k\, \sigma + k\, dk\, \sigma \ .
\label{YMCStransformations}
 \eea
 The relation (\ref{brstCSYM}) shows that the YM BRST operator $\delta_{YM}$ has the same
 algebraic content as the CS BRST operator $\delta$. As a matter of fact one sees from (\ref{YMcontact})
 that $\delta_{YM}$ differs from the CS $\delta$ only because it  also includes, on top of the CS transformations,
  the shift symmetry    
 \bea
 A \to A + k \, \zeta
 \label{YMshift}
 \eea 
together with $\sigma$, the BRST partner of $\zeta$. This shift symmetry  was originally  introduced in \cite{Beasley:2005vf}, to explain localization of CS theory on Seifert manifolds. 

Since $\zeta, \sigma$ make a trivial BRST doublet, the physical content of $\delta_{YM}$  and $\delta$ 
is the same. Indeed, from (\ref{YMCStransformations}) one derives the identity
\bea
S_{YM}\, \Gamma[\mathcal{A}^\prime(\mathcal{A})] = S_{CS}\,\Gamma[\mathcal{A}] \Big|_{\mathcal{A}\to \mathcal{A}^\prime(\mathcal{A})} \ .
\eea
Thus,  given any action $\Gamma[\mathcal{A}]$ invariant under the CS BRST operator $S_{CS}$, we can obtain an action invariant
under  $S_{YM}$ by performing the substitutions
\bea
c^\prime &=& c \ ,
\qquad A^\prime = A+ k\,\sigma \ ,
\qquad\tilde{A}^\prime =  \tilde{A} \ ,
\qquad \tilde{c}^\prime =  \tilde{c} +k\, dk \,\sigma \ .
\eea
In particular from the $S_{CS}$-invariant action  (\ref{actionCSbis}), one obtains the $S_{YM}$ invariant action
\bea
&&\tilde{\Gamma}_{CS}[A, \tilde{A}, \sigma]= \Gamma_{CS}[A +k\, \sigma] + \frac{1}{ 2}\,\int_{M_3} \Tr\, i_\gamma(\tilde{A})\, \tilde{A} \ ,
\label{actionCSbistilde}
\eea
which is equivalent to the CS action coupled to topological gravity backgrounds. This action is essentially the same as the Beasley-Witten's action \cite{Beasley:2005vf}. In the symplectic formalism of \cite{Beasley:2005vf} the term quadratic in $\tilde{A}$ is interpreted  as the symplectic 2-form on the space of
connections living on the base of the Seifert fibration. In our approach this term emerges naturally
from the coupling to the topological backgrounds. 

Summarizing, it is possible to include a CS-term in the action for the topological YM gauge theory coupled
to topological gravity: this is given in (\ref{actionCSbistilde}). The total action of the topological
YM+CS system coupled to the topological gravity backgrounds has therefore the form
\bea
\label{eq:YMCSactionfinal}
\Gamma_{YM+CS+t.g.} = S_{YM}\, \chi+\Gamma_{CS}[A +k\, \sigma] + \frac{1}{ 2}\,\int_{M_3} \Tr\, i_\gamma(\tilde{A})\, \tilde{A} \ .
\label{YMCSaction}
\eea

\section{The supersymmetric point}
The quantum partition function of the topological YM + CS system in presence of  topological gravity backgrounds\footnote{The notation in (\ref{CSPartitionFunction}) is schematic: we did not include explicitly the ghost sector which fixes the standard YM gauge invariance.}
\bea
Z[g_{\mu\nu}, \psi_{\mu\nu}, \gamma^\mu] = \int [dA \,d\tilde{A}\,d\tilde{c}\, d\sigma\,d\zeta] \, \mathrm{e}^{-\Gamma_{YM+CS+t.g.}} \ ,
\label{CSPartitionFunction}
\eea
 is an {\it equivariant} functional of the topological gravity multiplet. This means that it is both independent of $\xi^\mu$ and invariant under reparametrizations.  At classical level it satisfies the Ward identity
 \bea
 s\, Z[g_{\mu\nu}, \psi_{\mu\nu}, \gamma^\mu] = S\, Z[g_{\mu\nu}, \psi_{\mu\nu}, \gamma^\mu] =0 \ ,
 \label{ClassicalCSWard}
 \eea
 which can be --- and actually is ---  broken by quantum anomalies. We postpone the discussion regarding  quantum topological anomalies to Section \ref{Section:seiferttopanomaly}.

We can now look for bosonic backgrounds which are left invariant by $S$: 
\bea
\bar{\psi}_{\mu\nu}=0 \ ,\qquad \mathcal{L}_{\bar{\gamma}} \, \bar{g}^{\mu\nu} \equiv\bar{D}^\mu\,\bar{\gamma}^\nu+ \bar{D}^\nu\,\bar{\gamma}^\mu=0 \ .
\label{killinggamma}
\eea
The second equation above says that  the superghost background $\bar{\gamma}^\mu(x)$ is a Killing vector of the 3-dimensional
metric $\bar{g}_{\mu\nu}(x)$.  These geometrical data define a so-called {\it Seifert} structure on a 3-dimensional manifold
\cite{Beasley:2005vf}.  We see therefore  that  supersymmetric YM+CS theories admitting  a rigid topological supersymmetry  are precisely those defined on 3-dimensional Seifert manifolds. Hence one anticipates that supersymmetric YM+CS theories on Seifert manifolds  enjoy localization properties \cite{Pestun:2007rz}.
This  fact, originally discovered in a ``phenomenological'' way in \cite{Rozansky:1993zx},
has been subsequently explained using various approaches by different authors 
\cite{Beasley:2005vf},\cite{Kapustin:2009kz}.
In our approach this follows straighforwardly from the BRST trasformations of topological gravity.

The  topological YM+CS action on a fixed Seifert manifold 
\bea
\bar{\Gamma} 
& =& S_{YM}\, \chi+\Gamma_{CS}[A +k\, \sigma] 
 +\int_{M_3} \!\!\!\Tr\, \bigl[ \frac{1}{ 2}\, i_{\bar\gamma}(\tilde{A})\, \tilde{A}+\Lambda\ast i_{\bar\gamma}( \ast \tilde{A})+ b\ast i_{\bar\gamma}( \ast \,i_{\bar\gamma}(\tilde{c}+k\, dk \,\sigma))+ \nn\\
&&\qquad\qquad  + b\ast i_{\bar\gamma}( \ast \,(F + D\,(k\,\sigma)))\bigr] \ ,
\eea
is therefore invariant under the following BRST transformations which encode both gauge-invariance and global topological supersymmetry 
\bea
&&S_{YM}\, c = -c^2 + i_{\bar\gamma}(A)+ \sigma\ ,\nn\\ 
&&S_{YM}\, A = -Dc +i_{\bar\gamma}(\tilde{A})+ k\, \zeta \ ,\nn\\
&&S_{YM}\, \tilde{A} =  -[c, \tilde{A}] - F +i_{\bar\gamma}(\tilde{c})+  k\, D\, \sigma \ ,\nn\\
&&S_{YM}\,\tilde{c} = -[c,\tilde{c}] - D\,\tilde{A}- [\sigma, \tilde{A}]+ k\,dk\, \zeta \ ,\nn\\
&& S_{YM}\, \sigma = -[c,\sigma]+ \zeta \ ,\nn\\
&& S_{YM}\, \zeta = -[c,\zeta]+ i_{\bar\gamma}(D\,\sigma) \ .
\label{brsmatterequivariantfixed}
\eea

\section{The relation between  topological and  physical supersymmetry}
  
 The rigid topological theory that we
  obtained by coupling topological YM+CS to topological gravity computes certain (semi)-topological observables
  of the ``physical''  globally supersymmetric YM+CS theory living on the same manifold. In particular
  the topological partition function, which is the object that we consider in this paper, is the same as the superpartition function of the ``physical'' theory, i.e. the partition function with supersymmetric boundary conditions
  on both bosons and fermions.  Indeed,  as we mentioned in the introduction, the almost totality of the computations
  of  those (semi)-topological observables performed in recent years, were developed directly in the context of the ``physical''  theory with spinorial  supercharges.  We argued in the introduction that the  topological gravity viewpoint provides some benefits, both conceptual and practical. For starters, we just saw in the previous section that the Seifert condition emerges from topological gravity directly  --- without the necessity to go through covariantly constant spinors \cite{Kapustin:2009kz},  add
  extra symmetries\cite{Beasley:2005vf}, or  pick up ingenious gauges \cite{Blau:2006gh}.  But, above
  all, the coupling to topological gravity will allow us,  in the next sections, 
  to compute the moduli dependence of the partition function of supersymmetric YM+CS theory (involving  only vector multiplets)   by solving the anomalous topological Ward identities, in a completely regularization
  and gauge independent way.
  
 In this section we will describe more precisely the relation between the topological YM+CS obtained via
 the coupling to topological gravity   and   ``physical'', spinorial,  supersymmetric theory. We will also elucidate
 how the action that emerges from topological gravity encompasses  the topological actions which were introduced in  either  \cite{Beasley:2005vf} or \cite{Kallen:2011ny}. 
   
Supersymmetric CS  (SCS) theory
on curved space has been studied starting from  \cite{Kapustin:2009kz}, who considered the special
example of $S^3$. The
supersymmetric extension of the CS action in flat space \cite{Schwarz:2004yj} writes
\bea
\Gamma_{SCS} = \Gamma_{CS}+\int\, d^3x\, \Tr\,\bigl(D\, \sigma -\frac{1}{2}\,  \lambda^\dagger\,\lambda\bigr) \ ,
\label{SCS}
\eea
where the scalars $D$, $\sigma$ and the Dirac spinor $\lambda$ are in the adjoint of the gauge group.   Since  $D$, $\sigma$ and  $\lambda$ are auxiliary non-dynamical
fields,  supersymmetric CS theory (\ref{SCS}) is physically equivalent to pure CS theory. The action  (\ref{SCS}) is invariant
under the {\it global} supersymmetry transformations which have the structure
\bea
\delta \equiv \delta_\epsilon + \delta_{\bar{\epsilon}} \ ,
\eea
where
\bea
&&\delta_\epsilon\, A_\mu =  -\frac{i}{2}\, \lambda^\dagger\,\gamma_\mu\, \epsilon \ ,\nn\\
&&\delta_\epsilon \, \sigma =- \frac{1}{2}\,\lambda^\dagger\, \epsilon \ ,\nn\\
&& \delta_\epsilon\, D= - \frac{i}{2}\, D_\mu\lambda^\dagger\,\gamma^\mu\,\epsilon  +\frac{i}{2}\, [\lambda^\dagger, \sigma]\,\epsilon \ ,
\nn\\
&&\delta_\epsilon\,\lambda = - i\,\sqrt{g}\,\epsilon_{\mu\nu\rho}\,\gamma^\rho\, F^{\mu\nu}-D\, \epsilon+i\,\gamma^\mu\, D_\mu\,\sigma\, \epsilon \ ,
\nn\\
&&\delta_\epsilon\,\lambda^\dagger =0 \ ,
\label{globalCSsusy}
\eea
and $ \delta_{\bar{\epsilon}}\, \Phi =  (\delta_\epsilon\,\Phi)^\dagger$, for any field $\Phi \in\{ A, \sigma, D, \lambda, \lambda^\dagger\}$.

The classical action of ordinary, non supersymmetric,  CS action has the peculiarity of being invariant under local diffeomorphisms without
the need of introducing a space-time metric. This means that one can study quantum CS theory
on any {\it fixed} curved manifold: topological invariance of CS theory must be thought of as a global symmetry,
in the sense that one does not need to integrate over space-time metrics to make sense of the quantum theory.
This global symmetry is actually broken by anomalies: but, precisely because one is  dealing with a global symmetry, this does not spoil the consistency of the quantum theory. 

One might imagine that, by analogy,  SCS theory might be made invariant under local supersymmetry transformations without the need of explicitly introducing supergravity backgrounds. If this were so, SCS could be formulated consistently on any manifolds. Let us discuss why this is not the case.

The standard recipe for putting a generic supersymmetric theory on a curved manifold is to first couple it to supergravity, by promoting the global
supersymmetry  transformations (\ref{globalCSsusy}) to local ones.  For the SCS theory, 
this would mean in principle  to couple  (\ref{SCS})  to the  Noether supercurrents  
\bea
S^\mu = \frac{i}{2}\, \lambda^\dagger\,\gamma^\mu \,\sigma \ , \qquad \bar{S}^\mu =- \frac{i}{2}\,\gamma^\mu\,\lambda\,\sigma \ ,
\label{CSsupercurrents}
\eea
by changing the action 
\bea
&& \Gamma^{\mathrm{curved}}_{SCS}= \Gamma_{SCS}+\psi_\mu^\dagger\, \bar{S}^\mu+ S^\mu\, \psi_\mu +\cdots \ ,\nn\\
\eea
where $\psi_\mu$ is the gravitino field and the dots denote the higher order terms of the Noether procedure. The coupled action $\Gamma^{\mathrm{curved}}_{SCS}$ is invariant --- at linearized level --- under local supersymmetry transformations (\ref{globalCSsusy}) of the fields if the gravitino background also transforms as
\bea
\delta_\epsilon\,\psi_\mu= D_\mu \epsilon +\cdots \ , \qquad \delta_{\bar\epsilon}\,\psi^\dagger_\mu= D_\mu \epsilon^\dagger+\cdots \ .
\eea
However SCS theory is ``almost'' topological.  This is reflected by the fact that the supercurrents (\ref{CSsupercurrents}) vanish on shell
\bea
&& S^\mu = i\, \alpha \, \frac{\delta\, \Gamma_{SCS}}{\delta\lambda}\gamma^\mu\,\sigma+  \frac{i}{2}\,(1-\alpha) \lambda^\dagger\,\gamma^\mu \,\frac{\delta\, \Gamma_{SCS}}{\delta D} \ ,
\eea
where $\alpha$ is an arbitrary parameter. Since, when $\epsilon$ is space-dependent, the supersymmetry variation of the flat space action  writes in terms of the supercurrents as follows
\bea
 \bigl(\delta_\epsilon +\delta_{\bar{\epsilon}}\bigr) \Gamma_{SCS} &=& \int d^3x\, \bigl( S^\mu\, D_\mu\,\epsilon +  D_\mu\,\epsilon^\dagger\,\bar{S}^\mu\bigr)=\nn\\
&=& \int d^3x \, \bigl(  i\, \alpha \, \frac{\delta\, \Gamma_{SCS}}{\delta\lambda}\gamma^\mu\,\sigma+  \frac{i}{2}\,(1-\alpha) \lambda^\dagger\,\gamma^\mu \,\frac{\delta\, \Gamma_{SCS}}{\delta D}\bigr)\, D_\mu\,\epsilon  + c.c. \ ,
\label{SCSnoether}
\eea
one sees there is an alternative way to make (the diffeomorphism invariant extension of) $ \Gamma_{SCS}$  {\it locally} supersymmetric. This alternative method does not require introducing the gravitino:  thanks to on-shell
vanishing of the supercurrents one can simply modify the supersymmetry variations of $\lambda, \lambda^\dagger$ and $D$
\bea
 &&\tilde{\delta}_\epsilon\,\lambda = \delta_\epsilon\,\lambda + i\, \alpha \,\gamma^\mu\, D_\mu\, \epsilon\, \sigma \ ,\nn\\
&& \tilde{\delta}_\epsilon\, D= \delta_\epsilon\, D- \frac{i}{2}\,(1-\alpha)\, \lambda^\dagger\,\gamma^\mu \,D_\mu\,\epsilon \ ,
\label{localCSsusy}
\eea
where the covariant derivatives are those appropriate to the chosen curved manifold. 
Then, (\ref{SCSnoether}) is obviously equivalent to
\bea
\tilde{\delta}_\epsilon\,\Gamma_{SCS}=\tilde{\delta}_{\bar\epsilon}\,\Gamma_{SCS} =0 \ ,
\eea
for space-time dependent  $\epsilon$'s. 

The trouble with this ``alternative'' way to make the supersymmetry local is that, for $\alpha$ arbitrary and for generic manifolds,  the local supersymmetry algebra does not close.  By analyzing the supersymmetry commutation relations one discovers \cite{Hama:2010av}
that closure of the algebra requires both that the condition
\bea
\alpha = \frac{2}{3}
\label{SUSYclose}
\eea
is met {\it and} that the space-time dependence of $\epsilon$ be restricted by the differential equation
\bea
\gamma^\mu\,\gamma^\nu\,D_\mu \,D_\nu\,\epsilon = h \,\epsilon \ .
\label{SUSYclosebis}
\eea
One concludes that CS theory with rigid supersymmetry can be constructed only on  manifolds for
which solutions of Eq. (\ref{SUSYclosebis}) exist. For those special manifolds one
can obtain the corresponding rigid supersymmetry transformations by replacing in (\ref{localCSsusy}) 
the spinors which solve Eq. (\ref{SUSYclosebis}). 

The lesson of this discussion is that, even for the ``almost topological'' supersymmetric CS theories one cannot
neglect the coupling of the (classically vanishing) supercurrents to the supergravity backgrounds. Indeed it has since  been understood \cite{Klare:2012gn},\cite{Closset:2012ru} that the conditions (\ref{SUSYclose}) and (\ref{SUSYclosebis})
are to be interpreted, in a model independent way, as the equations for the  vanishing of the supersymmetry variation of the gravitino background 
\bea
\delta_\epsilon\,\psi_\mu=\delta_{\bar\epsilon}\,\psi^\dagger_\mu=0 \ .
\label{globalSUSY}
\eea
The nice feature of  Eqs. (\ref{globalSUSY}),
which are  readily seen to imply (\ref{SUSYclosebis}), is their universality: they do not depend on the specific
theory one is considering and  they characterize manifolds on which field theories with global supersymmetry may
be constructed. The specific form of supersymmetry does instead depend on both the solution of (\ref{globalSUSY})
and the form of the coupling of the supergravity multiplet to the theory at hand.

It can be shown that in the  {\it real} case when a solution $\epsilon$ of $\delta_\epsilon\,\psi_\mu=0$
defines by conjugation a solution of $\delta_{\bar\epsilon}\,\psi^\dagger_\mu=0$, the vector 
\bea
\bar{\gamma}^\mu = \epsilon^\dagger\,\Gamma^\mu \, \epsilon
\eea
is a (real) Killing vector of the underlying 3-manifold
\bea
D_\mu\, \bar{\gamma}_\nu +D_\nu\, \bar{\gamma}_\mu =0 \ .
\label{SeifertKilling}
\eea
This explains in particular why CS theories on 3-manifolds admitting a U(1)  action --- known as Seifert 3-manifolds --- enjoy the localization property which was originally discovered in \cite{Rozansky:1993zx} in an experimental way. 
We have seen that in our topological approach, the Seifert condition (\ref{SeifertKilling}) emerged directly from
the topological gravity BRST transformation laws, with no reference to (generalized) covariantly constant spinors.
 
We can consider also the  supersymmetric YM action in the SCS theory:
\bea
\Gamma_{SCS+SYM}= \Gamma_{SCS}+ \Gamma_{SYM} \ ,
\eea
where\footnote{The parameter $r$ which appears in this formula is the radius of the $S^1$ of the Seifert fibration.} 
\bea
\Gamma_{SYM}=\mathrm{Tr} \big[\frac{1}{4} \,F_{\mu\nu}^2 +\frac{1}{2}\, D_\mu\sigma\,D^\mu\,\sigma + \frac{1}{2}\, (D+\frac{\sigma}{r})^2 + \frac{i}{2}\,\lambda^\dagger \gamma^\mu D_\mu \lambda + \frac{i}{2}\,\lambda^\dagger\, [\sigma, \lambda]- \frac{1}{4\, r}\, \lambda^\dagger\,\lambda\bigr] \ .
\eea
This  latter action is not only  invariant under the  global supersymmetry transformations (\ref{globalCSsusy}), (\ref{localCSsusy}), (\ref{globalSUSY}), but also supersymmetric exact
\bea
\epsilon^\dagger\,\epsilon \,\mathcal{L}_{YM}= \tilde{\delta}_\epsilon\,\tilde{\delta}_{\bar\epsilon}\, \mathrm{Tr} \,\frac{1}{4} \,\bigl(\lambda^\dagger\,\lambda - 2\, D\,\sigma\bigr) \ .
\eea
Therefore, cohomologically, the SCS+SYM system is equivalent to SCS theory. 

The supersymmetric SCS+SYM on a {\it fixed} Seifert manifold can be twisted to give a model with
a topological rigid symmetry.  This was done in \cite{Kallen:2011ny}.  
The physical supersymmetric vector multiplet (\ref{globalCSsusy}) includes a Dirac  fermion $\lambda$ which has 4 real components.  After the twist of \cite{Kallen:2011ny},  three of those  form the topological gaugino $\Psi$. This,
together with the scalar $\sigma$ and the gauge connection $A$ form the multiplet of topological
YM. The twisted supersymmetry transformations of this supermultiplet turn out to have the form (\ref{brsYMtgcomponents}), in which the topological gravity field $\gamma^\mu$ is replaced by  the Reeb vector field $\bar{\gamma}^\mu$.
 The remaining fermion gives rise to a scalar $\alpha$ of ghost number +1 which form, together with 
  the auxiliary scalar field $\mathrm{D} $ of ghost number +2, an additional BRST {\it trivial} doublet,
\bea
S_{YM}\,\alpha &=&  - [c, \alpha]+ \mathrm{D} + X(A,\sigma) \ ,\nn\\
S_{YM} \, \mathrm{D}  &=& -[c, \mathrm{D}] + i_{\bar{\gamma}}\, D\,\alpha + [\sigma, \alpha] - \tilde{S}\, X(A,\sigma) \ .
\label{kallenfour}
\eea
Note that $S_{YM}$ is nilpotent for any choice of the scalar function $X(A,\sigma)$ of ghost number 2.  The twist of the physical supersymmetric theory gives rise to a specific choice for $X(A,\sigma)$. We keep it arbitrary for the
moment, to better explain the connections with other approaches.

The additional BRST trivial doublet $(\alpha, \mathrm{D})$ can function as an antighost-Lagrange multiplier pair, by adding to the $S_{YM}$ invariant action (\ref{actionCSbistilde}) an  $S_{YM}$-trivial term
\bea
\Gamma^\prime_{CS} =  \Gamma_{CS}[A +k\, \sigma] + \frac{1}{ 2}\,\int_{M_3} k\, \Tr\, \Psi \, \Psi
 + S_{YM}\, \int_{M_3} k\, dk\, \Tr\, \alpha\, \sigma \ .
 \label{actiontildegf}
\eea
One can pick 
\bea
X(A,\sigma)=0 \ .
\eea
With this choice the gauge-fixing term in (\ref{actiontildegf}) fixes the Beasley-Witten shift-symmetry:  integrating out $\mathrm{D}$ puts $\sigma$ to zero and integrating out $\alpha$ sets $\zeta=0$ and thus $\Psi= i_\gamma(\tilde{A})$. We recover in this way our original CS action (\ref{actionCSbis}). 

The twist of the physical SCS action (\ref{SCS}) discussed in  \cite{Kallen:2011ny} gives instead the 
\bea
X(A,\sigma)= \frac{k\,F}{k\, dk} +\sigma = i_\gamma(\ast\, F) +\sigma \ .
\label{kallenchoice}
\eea
Note that the choice of $X(A, \sigma)$ introduces a {\it spurious} dependence of the BRST operator on the
metric compatible with the vector field $\bar{\gamma}^\mu$ which defines the Seifert structure. This dependence should of course drop out in physical observables, but this is not explicit in the framework of  \cite{Kallen:2011ny}.
The reason of course is that twisting a physical supersymmetric action corresponds to make a specific
choice for the gauge-fixing term of the topological action. This, although sometimes convenient to perform explicit computations,  leads to gauge-dependent BRST transformations, somehow obscuring the geometric content of the topological symmetry.  One appealing feature of  our treatment is that it makes manifest that the theory only depends on the Seifert structure.

\section{Topological Anomaly for Seifert manifolds}
\label{Section:seiferttopanomaly}

The classical Ward identity (\ref{ClassicalCSWard}) can be broken by quantum anomalies
 \bea
S\, \log Z[g_{\mu\nu}, \psi_{\mu\nu}, \gamma^\mu] \equiv S\, \int_{M_3} i\,\Gamma[g_{\mu\nu}]= \frac{c}{6}\, \int_{M_3} i\,  A^{(3)}_1[g_{\mu\nu}, \psi_{\mu\nu}] \ .
 \label{QuantumCSWard}
 \eea
where $c$ is the anomaly coefficient.  
 The topological anomaly describes therefore the response of the {\it quantum action density} $\Gamma[ g_{\mu\nu}]$ of the YM+CS topological system under a {\it generic} variation of the metric  $\delta g_{\mu\nu}\equiv \psi_{\mu\nu}$
\bea
\delta  \int_{M_3}\Gamma[ g_{\mu\nu}]  =   \frac{c}{6} \int_{M_3}A^{(3)}_1[g_{\mu\nu}, \delta g_{\mu\nu}] \ .
\eea
The topological anomaly 3-form $A^{(3)}_1$ is a local cohomology class of ghost number +1 of the
BRST operator of topological gravity, which must satisfy the Wess-Zumino consistency condition
\bea
S\,   A^{(3)}_1[g_{\mu\nu}, \psi_{\mu\nu}]= - d\,   A^{(2)}_2[g_{\mu\nu}, \psi_{\mu\nu}, \gamma^\mu]\ .
\eea
Topological anomalies in any dimensions were classified in \cite{Imbimbo:2009dy}.
In 3 dimensions we have a single representative of ghost number +1
\bea
  A^{(3)}_1[g_{\mu\nu}, \psi_{\mu\nu}]=  \epsilon^{\mu\nu\rho}\, R^\alpha_\mu
\, D_\nu\,\psi_{\rho\alpha}\, d^3x \ ,
\eea

From the structure of the anomaly, it is clear that the parity invariant topological YM part of the action
cannot contribute to $c$.  A non-trivial $c$ can only come from the CS  part $\Gamma_{CS+t.g.}$
of the action. Since this theory is equivalent to bosonic CS, we conclude that $c$ is nothing but
the coefficient of the framing anomaly of pure bosonic CS. For $SU(N)$ gauge theories this has been
computed in \cite{Witten:1988hf}\footnote{In this Section we introduced a factor $\frac{k}{2\,\pi}$ in the normalization
of the Chern-Simons action, to make it  gauge-invariant modulo $ 2\,\pi$. Notice also that the parameter $k$ which appears in this and in the following formulas should be taken as the  ``shifted'' $l+2$, with respect to the level $l$ of the Kac-Moody current algebra whose conformal blocks
map to the states obtained from canonically quantizing the theory.
}
:
\bea
c_{SU(N)} =\frac{1}{4\,\pi} \frac{N^2-1}{k}
\eea

3-dimensional diffeomorphisms are not anomalous.  Hence, there exists a renormalization prescription which gives rise to  an effective  (non-local) action $\Gamma[g_{\mu\nu}]$ which transforms as a 3-form  under 3-dimensional {\it generic} diffeomorphisms. To express this condition, it is useful to introduce 
the Bardeen-Zumino BRST operator $S_{\mathrm{diff}}$ \cite{Bardeen:1984pm} associated to 3-dimensional
diffeomorphisms:

\bea
S_{\mathrm{diff}} = \mathcal{L}_\xi - \{i_\xi, d\} \ ,\qquad S_{\mathrm{diff}}^2=0 \ ,
\eea
where $\xi= \xi^\mu\,\partial_\mu$ is the reparametrization ghost in 3-dimensions, and $\mathcal{L}_\xi$ denotes the action of the Lie derivative along $\xi$ on the metric $g_{\mu\nu}$. The equation 
\bea
S_{\mathrm{diff}} \, \Gamma[g_{\mu\nu}]=0 \ ,
\label{3diffeoinvariance}
\eea
precisely expresses the fact that the  quantum action density $\Gamma[g_{\mu\nu}]$ transforms as 3-form under re\-pa\-ra\-me\-tri\-za\-tions.

After these preliminaries,  let us  now make our main observation:  When considering YM+CS topological theories on Seifert manifolds, one relaxes the request (\ref{3diffeoinvariance}) of full 3-dimensional
reparametrization invariance. One is satisfied with invariance under reparametrizations which preserve the Seifert structure:
these are reparametrizations whose ghost fields $\xi^\mu$ commute with the Reeb vector $\bar{\gamma}^\mu$. 
Let us denote by
\bea
S^{Seif}_{\mathrm{diff}} =   \mathcal{L}_{\xi_{Seif}} - \{ i_{\xi_{Seif}}, d \} \ , \qquad   \mathcal{L}_{\xi_{Seif}}\,\bar{\gamma} =0 \ ,
\eea
the Bardeen-Zumino BRST operator associated to diffeomorphisms preserving $\bar{\gamma}^\mu$. One also restricts the topological gravity background fields to those left  invariant under 
$\mathcal{L}_{\bar{\gamma}}$
\bea
\mathcal{L}_{\bar{\gamma}} \, g_{\mu\nu}= \mathcal{L}_{\bar{\gamma}} \, \psi_{\mu\nu}= \mathcal{L}_{\bar{\gamma}} \,\gamma^\mu=0 \ .
\label{seifertinvarianttg}
\eea
To parametrize solutions of (\ref{seifertinvarianttg}) it is useful  to introduce systems of coordinates adapted to the Seifert structure associated to $\bar{\gamma}^\mu$:
\bea
&& (ds)^2_M = \mathrm{e}^\sigma\, k\otimes k + g_{ij}\, dx^i\otimes\, dx^j =\nn\\
&&\qquad = \mathrm{e}^\sigma\,dy\otimes dy + 2\,  \mathrm{e}^\sigma\,a_i\, dx^i\otimes dy + (g_{ij} + \mathrm{e}^\sigma\, a_i\, a_j)\, dx^i\otimes dx^j \ ,
\label{adaptedmetric}
\eea
where $k$ is the contact  1-form
\bea
k \equiv  dy + a_i\, dx^i \ ,
\eea
dual to the Reeb vector field
\bea
i_{\bar{\gamma}}(k) =1 \ ,
\eea
$\sigma$, $g_{ij}$ and $a_i$ are fields on the two-dimensional surface $\Sigma_2$, associated to the Seifert fibration
$\pi: M\to \Sigma_2$.  The invariant $\psi_{\mu\nu}$ are analogously parametrized by fermions $\zeta, \psi_{ij}$ and $\psi_{i}$ 
living on $\Sigma_2$,  defined as follows:
\bea
\psi_{\mu\nu} = \begin{pmatrix}\mathrm{e}^\sigma \,\zeta & \mathrm{e}^\sigma \,\psi_i +  \mathrm{e}^\sigma \,\zeta\,a_i \\
\mathrm{e}^\sigma \,\psi_i +  \mathrm{e}^\sigma \,\zeta\,a_i & \psi_{ij} +  \mathrm{e}^\sigma \,(\psi_i\, a_j + \psi_j\, a_i + \zeta \, a_i\, a_j) 
\end{pmatrix} \ .
\label{Seifertgravitino}
\eea
Finally, the invariant $\xi^\mu$ and $\gamma^\mu$ can be written in terms of fields living on  $\Sigma_2$ as
\bea
\xi^\mu = (\xi^0, \xi^i)\equiv  (\xi^0, \vec{\xi}) \ ,\qquad \gamma^\mu = (\gamma^0, \gamma^i) \equiv (\gamma^0, \vec{\gamma}) \ .
\eea

Therefore, in the Seifert case,  the effective action $ \Gamma^{Seif}[ g_{ij}, \sigma, a_i] $ 
is a functional of the fields $\sigma$, $g_{ij}$ and $a_i$, and the appropriate renormalization prescription 
writes
\bea
S^{Seif}_{\mathrm{diff}} \, \Gamma^{Seif}[ g_{ij}, \sigma, a_i] =0 \ .
\label{Seifertprescription}
\eea

The action  density $\Gamma[g_{\mu\nu}]$ which satisfies the (strong) prescription (\ref{3diffeoinvariance})  
defines, of course, once written in Seifert adapted coordinates, also an action density
$ \Gamma^{Seif}[ g_{ij}, \sigma, a_i]$ satisfiying the (weaker)  Seifert renormalization condition (\ref{Seifertprescription}):
\bea
\Gamma^{Seif}[ g_{ij}, \sigma, a_i]= \Gamma[g_{\mu\nu}]\big|_{\mathcal{L}_{\bar{\gamma}} g_{\mu\nu}=0}
\label{wrongSeifertaction} \ .
\eea
This effective action satisfies the topological anomaly equation
\bea
s_{Seif} \, \int_{M_3}\Gamma^{Seif}[ g_{ij}, \sigma, a_i] =    \frac{c}{6} \int_{M_3} A^{(3)}_1[ g_{ij}, \sigma, a_i; \psi_{ij}, \zeta, \psi_i] \ ,
\eea
where the r.h.s. is obtained from $A^{(3)}_1[g_{\mu\nu}, \psi_{\mu\nu}]$ by evaluating it for $\bar{\gamma}$-invariant fields (\ref{seifertinvarianttg}).   
The BRST operator $s_{Seif}$ in the l.h.s. of the equation above  encodes topological gravity transformations which preserves the Seifert structure 
\bea
 &\!\! s_{Seif}\, \xi^i =  -\frac{1}{2}\, \mathcal{L}_{\vec{\xi}}\, \xi^i + \gamma^i \ , & s_{Seif}\, \gamma^i =  - \mathcal{L}_{\vec{\xi}}\,  \gamma^i \ ,\nn\\
& s_{Seif}\ g_{ij} = -\mathcal{L}_{\vec{\xi}}\, g_{ij} + \psi_{ij} \ , & s_{Seif}\, \psi_{ij} = -\mathcal{L}_{\vec{\xi}}\,  \psi_{ij}+ \mathcal{L}_{\vec{\gamma}} \, g_{ij} \ ,\nn\\
& \!\!\!\!s_{Seif}\, \xi^0  = -\mathcal{L}_{\vec{\xi}}\, \xi^0 + \gamma^0 \ , & s_{Seif}\, \gamma^0 = -\mathcal{L}_{\vec{\xi}}\,\gamma^0+\mathcal{L}_{\vec{\gamma}}\, \xi^0  \ ,\nn\\
 & s_{Seif}\, a_i  =  -\mathcal{L}_{\vec{\xi}}\,  a_i - \partial_i \xi^0 + \psi_i \ , & s_{Seif}\, \psi_i =  -\mathcal{L}_{\vec{\xi}}\,  \psi_{i} + \partial_i \gamma^0 + \mathcal{L}_{\vec{\gamma}}\, a_i \ ,\nn\\
 &\!\!\!\!\!\! \!\!s_{Seif}\, \sigma =  -\mathcal{L}_{\vec{\xi}}\,\sigma + \zeta \ ,
 & s_{Seif}\, \zeta =  -\mathcal{L}_{\vec{\xi}}\,\zeta +\mathcal{L}_{\vec{\gamma}}\, \sigma \ .
\label{SeifertBRST}
\eea
The invariant gravitational background fields split into three multiplets: one is the 2-dimensional topological gravity multiplet
$(\xi^i, \gamma^i, g_{ij}, \psi_{ij})$. Then there is an abelian topological gauge multiplet $(\xi^0,\gamma^0, a_i\,\psi_i)$:  their BRST properties are not just the ``flat'' ones, but they are modified by the coupling to 2-dimensional
gravity. Finally there is also an uncharged scalar topological multiplet $(\sigma, \zeta)$: this too is coupled to 2-dimensional topological gravity. 

Writing $A^{(3)}_1[ g_{ij}, \sigma, a_i; \psi_{ij}, \zeta, \psi_i]$ in adapted coordinates
\bea
A^{(3)}_1[ g_{ij}, \sigma, a_i; \psi_{ij}, \zeta, \psi_i]= \frac{1}{2} \mathcal{A} \, dy\,\epsilon_{ij} dx^i\,dx^j \ ,
\eea
one finds the following expression for  $\mathcal{A}$:
\bea
&&\mathcal{A} = -\frac{1}{2}\, \sqrt{g}\,\psi_{ij}\, \mathrm{e}^{\sigma}\, \bigl[D^i\, D^j\, f+ 3\, D^i f\, D^j\,\sigma+ 2\, f\, D^i\sigma\, D^j\sigma+f\, D^i\, D^j\sigma\bigr]+\nn\\
&&\qquad +\frac{1}{2}\,\sqrt{g}\, \mathrm{e}^{\sigma}\, \psi^i_i\, \Bigl[ D^2\,f+f\, (\frac{1}{2}\,R_2 - \mathrm{e}^{\sigma}\,f^2)+ \nn\\
&&\qquad +\frac{5}{2}\, D_j f\, D^j\sigma +\frac{3}{2}\, f\, D_j\sigma\, D^j\sigma+ \frac{3}{2}\, f\, D^2\, \sigma\Bigr]+\nn\\
&&\qquad + \frac{1}{2}\,\epsilon^{ij}\, \mathrm{e}^{\sigma}\, \psi_i\, \Bigl[ 6\,  \mathrm{e}^{\sigma}\, f\,D_j f- D_jR_2+\nn\\
&&\qquad + 6\, \mathrm{e}^{\sigma}\, f^2\, D_j \sigma- R_2 \, D_j\sigma- D_j\, D^2\,\sigma- D_j\sigma\, D^2\,\sigma\Bigr]+\nn\\
&&\qquad +\sqrt{g}\, \mathrm{e}^{\sigma}\, \zeta\,\Bigl[ \mathrm{e}^{\sigma}\, f^3- \frac{1}{2}\,f\, R_2- D_i\,f\, D^i\sigma- \frac{1}{2}\, f\,D_i\sigma\, D^i\sigma- f\, D^2\sigma- \frac{1}{2}\, D^2\, f\Bigr] \ .
\label{anomaly2dextended}
\eea
In (\ref{anomaly2dextended}), $R_2$ is the scalar curvature constructed with the 2-dimensional metric $g_{ij}$ and 
\bea
f =  \frac{\epsilon^{ij}}{\sqrt{g}}\,f_{ij} =\frac{\epsilon^{ij}}{\sqrt{g}}\, (\partial_i\,a_j-\partial_j\,a_i) \ ,
\eea
is the scalar field dual to the $U(1)$ field strength $ f^{(2)}\equiv d a$. 

The important fact, now,  is that $ A^{(3)}_1[ g_{ij}, \sigma, a_i; \psi_{ij}, \zeta, \psi_i]$ is $s_{Seif}$-trivial\bea
  \mathcal{A} =s_{Seif} \, \Gamma^{Seif}_{WZ}[ g_{ij}, \sigma, a_i] \ ,
\eea
where the Wess-Zumino action $\Gamma^{Seif}_{WZ}$ is the following {\it local} functional \footnote{The topological anomaly $A^{(3)}_1$ satisfies also: $A^{(3)}_1= S\, \Gamma^{(3)}_{GCS}[g]$, where $\Gamma_{GCS}[g]$ is the gravitational Chern-Simons action. 
In \cite{Beasley:2005vf} it is pointed out that on a Seifert manifold $M$ there exists a natural  trivialization of the double tangent $TM\oplus TM$ which leads to a specific definition of the Chern-Simons invariant $\Gamma_{GCS}[g]$. This provides a geometric explanation of why the topological anomaly is BRST trivial in the Seifert context. To the best of our knowledge however an explicit local Wess-Zumino functional of the adapted Seifert metric as $\Gamma^{Seif}_{WZ}[ g_{ij}, \sigma, a_i]$ has not
been presented earlier.  Note that $\Gamma^{(3)}_{GCS}[g]$ evaluated for an adapted Seifert metric  and $\Gamma^{Seif}_{WZ}[ g_{ij}, \sigma, a_i]$ differ by a not-globally defined total derivative, whose precise form is described in Appendix \ref{App:AppendixCGSWZ}.}

\bea
\Gamma_{WZ}^{Seif}[ g_{ij}, \sigma, a_i] =\frac{1}{2}\,\sqrt{g} \, \mathrm{e}^{2\,\sigma}\, f^3 - \sqrt{g}\,\frac{1}{2}\, \mathrm{e}^{\sigma}\,  f\, R_2- \frac{1}{2}\,\sqrt{g}\,  \mathrm{e}^{\sigma}\,f\, D^2\,\sigma \ .
\label{WZSeifertaction}
\eea
$\Gamma^{Seif}_{WZ}[ g_{ij}, \sigma, a_i]$ is a legitimate Wess-Zumino action since it is both {\it local} and invariant under re\-pa\-ra\-me\-tri\-za\-tions   which preserve the Seifert structure
\bea
S^{Seif}_{\mathrm{diff}} \,\Gamma^{Seif}_{WZ}[ g_{ij}, \sigma, a_i]=0 \ .
\eea
It should be kept in mind, however, that  $\Gamma^{Seif}_{WZ}[ g_{ij}, \sigma, a_i]$  --- unlike  the non-local $\Gamma[ g_{\mu\nu}]$ in eq. (\ref{wrongSeifertaction}) ---  is  {\it not} invariant under the full 3-dimensional $S_{\mathrm{diff}}$.  

Hence one can define the  effective action
\bea
 \tilde{\Gamma}^{Seif}[ g_{ij}, \sigma, a_i] \equiv  \Gamma^{Seif}[ g_{ij}, \sigma, a_i] - \frac{c}{6}\,\Gamma^{Seif}_{WZ}[ g_{ij}, \sigma, a_i] \ ,
 \label{correctSeifertaction}
\eea
which is both $s_{Seif}$-invariant --- i.e. topological in the Seifert sense --- 
\bea
s_{Seif}\,  \tilde{\Gamma}^{Seif}[ g_{ij}, \sigma, a_i]=0 \ ,
\label{eq:Gammatildesinv}
\eea
{\it and}  invariant under  reparametrizations which preserve $\bar{\gamma}$
\bea
S^{Seif}_{\mathrm{diff}} \,\tilde{\Gamma}^{Seif}[ g_{ij}, \sigma, a_i]=0
\label{eq:Gammatildeseifdiff} \ ,
\eea

Summarizing, we have shown that it is always possible to define through Eq. (\ref{correctSeifertaction})  a quantum action density $\tilde{\Gamma}^{Seif}[ g_{ij}, \sigma, a_i]$ which depends on the moduli parametrizing the Seifert structures (which we will characterize in Section \ref {Section:moduli}) but not on the 
specific adapted metric which one picks  to quantize the theory.  

\section{Moduli}
\label{Section:moduli}

In this Section we will identify cohomologically the parameters on which the quantum partition function of 3d supersymmetric gauge theory depends.  

In the general  ``non-real'' case  \cite{Closset:2012ru}, the space of infinitesimal deformations
of a given supersymmetric background can be identified \cite{Closset:2013vra} with the appropriate cohomology of a certain differential
operator $\tilde{\partial}$ which, loosely speaking, is the 3d generalization of the 2-dimensional Dolbeault operator $\bar{\partial}$. The cohomology of $\tilde{\partial}$ has not been well studied yet, and, to our knowledge, it is not explicitly known for general 3-manifolds. On the other hand, in the more special ``real'' context to which our analysis is confined one can give a very explicit and general characterization of  the space of infinitesimal deformations of (topological) supersymmetric backgrounds.  

We should also emphasize that the analysis of topological quantum anomalies contained in the previous Section makes our identification of the parameters which affect the partition function valid both at the non-linear and the quantum level. 

We have seen that supersymmetric topological backgrounds correspond to solutions of the Killing equations
\bea
\mathcal{L}_{\bar{\gamma}} \, \bar{g}_{\mu\nu} \equiv\bar{D}_\mu\,\bar{\gamma}_\nu+ \bar{D}_\nu\,\bar{\gamma}_\mu=0 \ .
\label{killinggammabis}
\eea
Given a solution $\{\bar{g}_{\mu\nu}, \bar{\gamma}^\mu\}$ of (\ref{killinggammabis}) we want explore nearby supersymmetric backgrounds $\{\bar{g}_{\mu\nu}+\delta g_{\mu\nu}, \gamma^\mu+\delta\gamma^\mu\}$.
The deformations $\{\delta g_{\mu\nu}, \delta\gamma^\mu\}$ must satisfy the linear equation
\bea
\mathcal{L}_{\bar{\gamma}} \,\delta g_{\mu\nu}+ \mathcal{L}_{\delta\gamma} \, \bar{g}_{\mu\nu} =0 \ .
\label{Seifertdeformations}
\eea
Let us  introduce the vector space 
\bea
V_0= \Gamma(TM_3)\oplus\mathrm{Sym}_2(TM_3) \ ,
\eea
where  $\Gamma(TM_3)$ is the space of vector fields on $M_3$ and $ \mathrm{Sym}_2(TM_3)$ the space of   2-index symmetric tensors on $M_3$. The deformation equation (\ref{Seifertdeformations})  describes
therefore the kernel of  the linear operator
\bea
&& Q_0: V_0\to V_1 \ ,\nn\\
&& Q_0: (\delta\gamma^\mu,\delta g_{\mu\nu})\to \mathcal{L}_{\bar{\gamma}} \,\delta g_{\mu\nu}+ \mathcal{L}_{\delta\gamma} \, \bar{g}_{\mu\nu} \ ,
\eea
where
\bea
V_1= \mathrm{Sym}_2(TM_3) \ .
\eea

We are interested in characterizing physical deformations,  i.e. solutions of this equation {\it modulo gauge equivalences}.  Gauge-invariance includes  infinitesimal diffeomorphisms
\bea
(\delta\gamma^\mu,\delta g_{\mu\nu})\sim (\delta\gamma^\mu,\delta g_{\mu\nu})+ (\mathcal{L}_\xi\delta\gamma^\mu,\mathcal{L}_\xi\delta g_{\mu\nu}) \ ,
\label{gaugeinvarianceone}
\eea
where $\xi^\mu$ is a vector field on $M_3$. But $\mathcal{L}_{\bar{\gamma}}$-invariant  topological deformations of the metric should also  be treated as a gauge invariances 
\bea
&&(\delta\gamma^\mu,\delta g_{\mu\nu})\sim (\delta\gamma^\mu,\delta g_{\mu\nu}+ \psi_{\mu\nu}) \ ,
\label{gaugeinvariancetwo}
\eea
for any $\mathcal{L}_{\bar{\gamma}}$-invariant  $\psi_{\mu\nu}$\footnote{The explicit form for invariant $\psi_{\mu\nu}$, in adapted coordinates, is given in Eq. (\ref{Seifertgravitino}).}
\bea
\psi_{\mu\nu}\in\mathrm{Sym}^{inv}_2(TM_3)\equiv \{ \psi_{\mu\nu}\in \mathrm{Sym}_2(TM_3)\big| \;\mathcal{L}_{\bar{\gamma}} \psi_{\mu\nu}=0  \} \ .
\eea

We can therefore define a linear operator $Q_{-1}$ which captures both gauge equivalences (\ref{gaugeinvarianceone}) and (\ref{gaugeinvariancetwo})
\bea
&& Q_{-1} : V_{-1}\to V_0 \ ,\nn\\
&& Q_{-1} :  \{ \xi^\mu,\psi_{\mu\nu}\}\to \{\mathcal{L}_\xi\bar{g}_{\mu\nu}+ \psi_{\mu\nu}, \mathcal{L}_\xi\,\bar{\gamma}^\mu\} \ ,
\eea
where
\bea
V_{-1}= \Gamma(TM_3)\oplus\mathrm{Sym}^{inv}_2(TM_3) \ .
\eea
We have
\bea
Q_0 \, Q_{-1}=0 \ .
\eea
One can consider therefore the complex
\bea
0  \rightarrow V_{-1} \xrightarrow{Q} V_0  \xrightarrow{Q}V_1 \rightarrow 0
\label{seifertsequence} \ .
\eea
The associated cohomology space
\bea
H_0(Q) = \frac{\ker Q_0}{\mathrm{Im}\, Q_{-1}} \ ,
\eea
describes therefore inequivalent deformations around  the Seifert structure $\{\bar{g}_{\mu\nu}, \bar{\gamma}^\mu\}$.
The kernel of $Q_{-1}$
\bea
\ker Q_{-1} =  H_{-1}(Q) =\{ (\xi^\mu,\psi_{\mu\nu}): [\xi, \bar{\gamma}]=0, \psi_{\mu\nu} = - \mathcal{L}_\xi\,\bar{g}_{\mu\nu}\} \ ,
\eea
is isomorphic to the commutant $C_{\bar{\gamma}}$ of $\bar{\gamma}^\mu$ in the Lie algebra of vectors
fields on $M_3$:
\bea
C_{\bar{\gamma}}= \{\gamma^\mu\in  \Gamma(TM_3): [\gamma, \bar{\gamma}]=0\}\simeq  H_{-1}(Q) \ .
\eea
Consider now the map between  $C_{\bar{\gamma}}$ and $\mathrm{Sym}^{inv}_2(TM_3)$  :
\bea
&& \varphi: C_{\bar{\gamma}}\to \mathrm{Sym}^{inv}_2(TM_3) \ ,\nn\\
&& \varphi: \gamma^\mu  \to \mathcal{L}_\gamma \bar{g}_{\mu\nu} \ .
\eea
The kernel of $\varphi$  is made of the isometries of $\bar{g}_{\mu\nu}$ which commute with $\bar{\gamma}^\mu$
\bea
\ker \varphi = \{ \gamma: [\gamma, \bar{\gamma}]=0, \; \mathcal{L}_\gamma \bar{g}_{\mu\nu}=0\}\subset C_{\bar{\gamma}} \ .
\eea
The cokernel of $\varphi$ is, on the other hand, characterized by the $ \mathcal{L}_{\bar{\gamma}}$-invariant $\psi_{\mu\nu}$'s which are orthogonal to $\mathrm{Img}\, \varphi$
\bea
 0= \int_{M_3}\gamma^\nu \bar{D}^\mu\,\psi_{\mu\nu}= \int_{M_3}\gamma^\nu  g_{\nu\mu}\,v^\mu \equiv \langle\gamma, v\rangle \qquad \forall \gamma \in C_{\bar{\gamma}} \ ,
 \label{cokernelphi}
\eea
where
\bea
v^\mu \equiv \bar{g}^{\mu\lambda}\bar{D}^\nu\,\psi_{\lambda\nu} \ .
\eea
The vector $v^\mu$ is   $ \mathcal{L}_{\bar{\gamma}}$-invariant, since $\bar{g}_{\mu\nu}$ and $\psi_{\mu\nu}$ are: 
\bea
 \mathcal{L}_{\bar{\gamma}} v^\mu =0= [\bar{\gamma},v] \ .
 \eea
Hence $v^\mu\in  C_{\bar{\gamma}}$. But since, according to (\ref{cokernelphi})  $v^\mu$ is orthogonal to whole
$C_{\bar{\gamma}}$, it vanishes
\bea
v^\mu= \bar{g}^{\mu\lambda}\bar{D}^\nu\,\psi_{\lambda\nu}=0 \ .
\eea
We conclude that 
\bea
\mathrm{coker}\,\varphi =\{ \psi_{\mu\nu}: \mathcal{L}_{\bar{\gamma}}\,\psi_{\mu\nu}=0, \bar{D}^\mu\,\psi_{\mu\nu}=0\} \ ,
\eea
 and therefore
 \bea
\mathrm{Sym}^{inv}_2(TM_3) = \mathrm{Im} \,\varphi \oplus\mathrm{coker}\, \varphi  \simeq \frac{C_{\bar{\gamma}}}{\ker\, \varphi} \oplus \mathrm{coker}\, \varphi \ .
\eea
Let us now consider the cokernel of $Q_0$: it  is characterized by the equation
\bea
\int_{M_3} \psi^{\mu\nu}\,( \mathcal{L}_{\bar{\gamma}} \,\delta g_{\mu\nu}+ \mathcal{L}_{\delta\gamma} \, \bar{g}_{\mu\nu})=0\qquad  \forall\; \delta\gamma^\mu\in \Gamma(TM_3)\; \mathrm{and}\;\forall\; \delta g_{\mu\nu}\in \mathrm{Sym}_2(TM_3) \ .
\eea
This implies
\bea
 \mathcal{L}_{\bar{\gamma}} \,\psi^{\mu\nu}=\bar{D}^\mu \,\psi_{\mu\nu}=0 \ .
\eea
In other words
\bea
\mathrm{coker}\, Q_0 = H_1(Q) =\mathrm{coker}\, \varphi \ .
\eea
The exactness of the sequence (\ref{seifertsequence}) implies therefore 
\bea
T_{\{\bar{g}, \bar{\gamma}\}}{\mathcal{M}} \simeq H_0(Q) = \frac{H_{-1}(Q)\oplus H_1(Q)}{\mathrm{Sym}^{inv}_2(TM_3)} \simeq \ker \varphi \ ,
\eea
where $T_{( \bar{\gamma}, \bar{g})}\mathcal{M}$ is the tangent to the space of physical moduli of the theory at a point 
$\{\bar{g}_{\mu\nu}, \bar{\gamma}^\mu\}$.

Hence $\bar{g}_{\mu\nu}$-isometries which commute with $\bar{\gamma}^\mu$ are in one-to-one correspondence with
non-trivial deformations  of a given Seifert structure $\{\bar{g}_{\mu\nu}, \bar{\gamma}^\mu\}$.
 $\bar{\gamma}^\mu$ itself, of course, is always one of such isometries.  The corresponding deformation is a rescaling
of $\bar{\gamma}^\mu$. Since a rescaling of $\bar{\gamma}^\mu$ in the YM + CS action (\ref{eq:YMCSactionfinal})
can be reasorbed in a rescaling of the field $\tilde{A}$, the YM + CS partition function does not depend on this kind
of deformation. In conclusion the parameter space,  which the YM + CS partition function on a Seifert manifold depends on, is  the quotient space
\bea
\mathrm{ker} \,\varphi/\sim \ ,
\eea
where the equivalence relation is
\bea
\gamma^\prime \sim \gamma + \alpha \bar{\gamma}, \qquad \gamma,\gamma^\prime \in \ker \varphi \ ,
\eea
with $\alpha$ constant.

\section{The topological anomaly for the squashed spheres}
 Let us now consider the squashed metric on $S_3$\footnote{Our definitions and conventions for the Hopf coordinates for the
 squashed sphere are reviewed in Appendix \ref{App:AppendixSquashed}.}:
 \bea
ds^2 = \bar{g}_{\mu\nu}(x; l,\tilde{l})\, dx^\mu\otimes dx^\nu=(l^2\,\sin^2\theta+ \tilde{l}^2\,\cos^2\theta)\, d\theta^2 + l^2\,\cos^2\theta\, d\phi_1^2 + \tilde{l}^2\,\sin^2\theta\, d\phi_2^2 \ ,
\label{squashedmetricone}
\eea
where $\phi_{1,2}\in [0, 2\,\pi]$ and $0\le \theta\le \frac{\pi}{2}$, are the Hopf coordinates on $S_3$.

The vector field
 \bea
 \bar{\gamma}^\mu\partial_\mu  =  \frac{1}{l}\,\partial_{\phi_1} + \frac{1}{\tilde{l}}\,\partial_{\phi_2}= \frac{\partial}{\partial y} \ ,
 \label{squashedreeb}
 \eea
 is, for each value of the  squashing parameters $(l, \tilde{l})$,  an isometry of $\bar{g}_{\mu\nu}(x; l,\tilde{l})$.  A system of coordinates $(y, \alpha, \beta)$ adapted to the Seifert structure  corresponding to $(l, \tilde{l})$  is defined by the relations
 \bea
  &&\theta= \frac{\alpha}{2} \ ,\qquad \phi_1= \frac{y}{l} + \frac{\beta}{2\,l}+\frac{\epsilon(\alpha,\beta)}{l} \ ,\qquad \phi_2= \frac{y}{\tilde{l}} -\frac{\beta}{2\,\tilde{l}}+\frac{\epsilon(\alpha,\beta)}{\tilde{l}} \ ,
   \eea
  and their inverse 
 \bea
&& y = \frac{l\, \phi_1+ \tilde{l}\, \phi_2}{2}-\epsilon(\alpha, \beta) \ ,\qquad
\beta = l\,\phi_1 - \tilde{l}\,\phi_2 \ ,\qquad \alpha = 2\,\theta \ .
\eea
$\epsilon(\alpha, \beta)$ is an arbitrary local function which corresponds to abelian gauge transformations
along the fiber of the fibration. The squashed metric (\ref{squashedmetricone}) writes in these adapted coordinates as follows
 \bea
  &&ds^2=(dy + a)^2 +\frac{1}{4}\, \bigl[ \frac{1}{2}\, (l^2+\tilde{l}^2 + (\tilde{l}^2-l^2)\,\cos \alpha)\,d\alpha^2+\sin ^2\alpha\, d\beta^2 \bigr] \ ,
\label{squashedmetricadapted}
  \eea
where the abelian gauge connection $a$ and its field strength $f^{(2)}$ are given by
  \bea
  &&a =\frac{1}{2}\, \cos \alpha\, d\beta+ d\,\epsilon \ , \nn\\
 &&f^{(2)}= da = -\frac{1}{2}\sin\alpha\, d\alpha\, d\beta \ .
   \eea
 The curvature $R_2$ of the 2-dimensional metric $g_{ij}$ on the $S_2$ base of the Seifert fibration is
  \bea
  && R_2 = \frac{8}{l^2}\, \frac{(b^4-1) \cos\alpha+2 \left(b^4+1\right)}{\left(\left(b^4-1\right) \cos\alpha+b^4+1\right)^2} \ ,\nn\\
  &&\sqrt{g} =l\,\frac{1}{4}\,\sin\alpha \sqrt{\frac{b^4-1}{2} \cos\alpha+\frac{b^4+1}{2}} \ ,
  \label{squashedcurvatures}
  \eea
  where we introduced  the ratio
  \bea
  b^2\equiv \frac{\tilde{l}}{l} \ .
  \eea 
  The scalar field which is dual to the abelian field strength is therefore
  \bea
  && f = \frac{\epsilon^{ij}\, \partial_i \, a_j}{\sqrt{g}}=\frac{2 \sqrt{2}}{l\,\sqrt{\left(b^4-1\right) \cos\alpha+b^4+1}} \ , \qquad \sqrt{g}\,f = \frac{1}{2}\, \sin\alpha \ .
  \eea
 
We learnt in the previous Section that, given  $\bar{g}_{\mu\nu}(x; l ,\tilde{l})$ and $\bar{\gamma} =  \frac{1}{l}\,\partial_{\phi_1} + \frac{1}{\tilde{l}}\,\partial_{\phi_2}$ , the deformations of the Seifert structure
are associated to the isometries which commute with $\bar{\gamma}$ modulo $\bar{\gamma}$. For generic $(l, \tilde{l})$
 the isometries which commute with $\bar{\gamma}$ are $\partial_{\phi_1}$ and $\partial_{\phi_2}$.  Hence we see that $b^2$ parametrizes precisely the inequivalent deformations of the Seifert
structure around a generic point $b\not =1$. The point $b=1$ corresponds to the ``round'' sphere,  which has an enhanced symmetry $SU(2)_L\times SU(2)_R$. Around this point more general deformations are possible, since the 
isometries which commute with, let us say, $J_3^R$ form the full $SU(2)_L$.
 
Let us  compute the topological anomaly for  the squashed sphere metric $\bar{g}_{\mu\nu}(x; l,\tilde{l})$.
Since $A^{(3)}_1[g_{\mu\nu}, \psi_{\mu\nu}]$ depends only on the conformal class of the metric, we can take, without loss of generality
\bea
l=1 \ ,\qquad \tilde{l}= b^2
\eea
and put  $\bar{g}_{\mu\nu}(x;1, b^2)\equiv \bar{g}_{\mu\nu}(x;b )$. Then
\bea
\bar{\psi}_{\mu\nu}(x; b) = b\,\partial_b\,\bar{g}_{\mu\nu}(x; b)= \begin{pmatrix} 4 \, b^4\, \cos^2\theta &0 &0  \\
0 & 0 & 0\\
0& 0 &  4\, b^4 \, \sin^2\theta \ ,
\end{pmatrix}
\eea 
where the $b$-derivative is taken by keeping the Hopf coordinates constant. It is easy to verify that the topological
anomaly for these backgrounds vanishes for all $b$'s:
\bea
 A^{(3)}_1[\bar{g}, \bar{\psi}] =0 \ .
 \label{vanishinganomalysquashed}
\eea
This implies that the effective action $\Gamma[g_{\mu\nu}]$ evalutated for the squashed sphere metric
$\bar{g}_{\mu\nu}(x;b )$ is independent of $b$:
\bea
b\,\partial_b  \Gamma[\bar{g}_{\mu\nu}(x;b)] = b\,\partial_b\,\Gamma^{Seif}[ \bar{g}_{ij}(X;b), \bar{\sigma}(X;b), \bar{a}_i(X;b)]=0 \ ,
\eea
where $X(x;b)\equiv (y, \alpha, \beta)$ are the coordinates adapted to the Seifert structure parametrized by $b^2$.

However,  we explained in Section \ref{Section:seiferttopanomaly} that $\Gamma[g_{\mu\nu}]$ is {\it not}
the action renormalized with the correct Seifert prescrition (\ref{Seifertprescription}).  The action  $\tilde{\Gamma}^{Seif}[ g_{ij}, \sigma, a_i] $ renormalized according to the Seifert prescription is given by (\ref{correctSeifertaction}). When this  latter action is evaluated on
$\bar{g}_{\mu\nu}(x;b )$, one obtains
\bea
 \tilde{\Gamma}^{Seif}[ \bar{g}_{ij}, \bar\sigma, \bar{a}_i] =  \Gamma^{Seif}[ \bar{g}_{ij}(X;b), \bar{\sigma}(X;b), \bar{a}_i(X;b)] - \frac{c}{6}\,\Gamma^{Seif}_{WZ}[ \bar{g}_{ij}(X;b), \bar{\sigma}(X;b), \bar{a}_i(X;b)].
\eea
We have  just shown that, due to  (\ref{vanishinganomalysquashed}), the first (non-local) term  in the r.h.s. of
the equation above is $b$-independent.  But the second (local) one is not. Indeed,   by plugging  (\ref{squashedcurvatures})  inside (\ref{WZSeifertaction}) one computes
 \bea
  &&  \Gamma^{Seif}_{WZ}[ \bar{g}_{ij}(X;b), \bar{\sigma}(X;b), \bar{a}_i(X;b)]= \int\frac{1}{2}\sqrt{g}\, f\, (f^2- R_2)= \nn\\
  &&\qquad =- (2\,\pi)^2\, \bigl( b^2+ \frac{1}{b^2}\bigr) \ ,
\eea
which therefore encodes the whole dependence of the Seifert partition function on $b^2$:
\bea
Z^{\mathrm{squashed}}(b) = \mathrm{e}^{ i\,\tilde{\Gamma}^{Seif}[ \bar{g}_{ij}, \bar\sigma, \bar{a}_i]}= \mathrm{e}^{i\,\frac{4\,\pi^2\,c}{3}\,[\frac{1}{2}\,( b^2+ \frac{1}{b^2})-1]}\, Z^{\mathrm{squashed}}(1)
\label{anomalyZsquashed} \ ,
\eea
where $Z^{\mathrm{squashed}}(1)$ is the partition function on the ``round'' sphere, corresponding to $b=1$.

Let us compare the anomaly equation (\ref{anomalyZsquashed}) with the YM+CS partition function on the squashed sphere computed directly  by means
of localization in \cite{Hama:2010av}, taking for simplicity the case of $SU(2)$ gauge group
\bea
 Z^{\mathrm{squashed}}(b)= \int_{-\infty}^{\infty} \frac{dx}{2\,\pi\,i} \, \mathrm{e}^{-\frac{i\, k\, x^2}{8\,\pi}}\, \sinh\frac{i\, x\, b}{2}\, \sinh\frac{i\, x}{2\,b}\ .
 \eea
Here the exponential in the integrand comes from the value of the action on the saddle point, while the 
hyperbolic sine factors are the results of the 1-loop determinants.  Expressing the $\sinh$ factors in terms
of exponentials, the integration reduces to the sum of Gaussian integrals:
\bea
 Z^{\mathrm{squashed}}(b)=\mathrm{e}^{\frac{i\,\pi}{2\, k}\, (b^2+\frac{1}{b^2})-\frac{i\,\pi}{4}}\, \, Z^{CFT}_{CS} \ ,
 \label{explicitZsquashed}
 \eea
 where
 \bea
 \, Z^{CFT}_{CS}= \sqrt{\frac{2}{k}}\,\sin\frac{\pi}{k}
 \eea
 is the round sphere $SU(2)$ CS partition function, obtained by means of surgery from the CFT modular transformations of the WZW 2d conformal model. The result (\ref{explicitZsquashed}) obtained by
 direct computation agrees with our prediction (\ref{anomalyZsquashed}) obtained from the topological anomaly and
 the Wess-Zumino Seifert action,  once we insert the topological anomaly  coefficient for $SU(2)$,
$c_{SU(2)} = \frac{1}{4\,\pi}\,\frac{3}{k}$.

\section{Conclusions} 

The current paradigm for localization relies on   spinorial  global supercharges. Since the fate
and properties of quantum global symmetries are best studied by introducing background fields coupled to currents, the same paradigm has lead to studying the coupling of ``physical'' supersymmetric  theories to off-shell supergravity.
In particular the search for globally supersymmetric models has been reduced to the study of generalized
covariantly constant spinors.

In this paper we proposed an alternative route.  Localization is naturally understood in terms of topological
scalar supercharges --- i.e. in terms of topological theories and BRST symmetries. In this framework it
is the coupling of topological field theories to {\it topological gravity}, not supergravity, which is relevant. 
For this reason we worked out the coupling  of both CS and topological
Yang Mills theory ---  i.e. of a generic vector twisted supermultiplet --- to topological gravity. 
 The BRST structure of the Chern-Simons supermultiplet  looks very different from that of the topological YM theory, when the latter is presented in its familiar formulation valid in arbitrary dimension. We exhibited, however,
 a new formulation of  the BRST transformations of topological YM in 3d  purely in  terms of the CS supermultiplet.
 This allowed us to derive a unique (anomalous) Ward identity which characterizes the coupling of a 3d generic twisted vector supermultiplet  to topological gravity. 
 
One first advantage of  the topological gravity viewpoint is that the structure of topological gravity is the same in all dimensions, a fact which makes the characterization of supersymmetric bosonic backgrounds straightforward. For example, in the context of 3d gauge theories which is the one of  this paper, we have seen that the Seifert condition emerges quite immediately without the need to go through generalized covariantly constant spinors or similar indirect routes peculiar to other approaches. Moreover we have found that, in the 3d context, the off-shell coupling of topological (YM+CS) gauge theories to topological gravity is easily achieved by suitably covariantizing the ``rigid'' coboundary  BRST operator with a universal term which is the form-contraction with the super-ghost field of topological gravity. We also discovered that the anti-fields of the BV formulations of CS theory are nothing but the auxiliary fields which are required to close off-shell the topological supersymmetry algebra.

But the real payoff of the topological approach was that it made straightforward to identify the subset
of local topological transformations which preserves  the Seifert backgrounds. These turned out to
be 2d topological gravity transformations coupled to topological abelian gauge transformations. 
This allowed us to give a cohomological characterization of the Seifert background moduli.  Moreover
we were able to explicitly solve the anomalous Ward identity associated to topological transformations of the gravitational
background. The solution involved a Wess-Zumino local action, invariant under the reparametrizations which preserve the Seifert structure.

The triviality of the topological (framing) 3d  anomaly when restricted on Seifert backgrounds shows, rigorously and in a completely regularization independent way,  that the quantum effective action of the gauge theory on Seifert manifold depends on the Seifert moduli  but not on the 
specific metric adapted to the Seifert structure which one picks  to quantize the theory.  
The Wess-Zumino Seifert action also completely determines the dependence of the partition function on the Seifert moduli. We explicitly showed this in the case of the squashed sphere, for which we recovered the dependence of the partition function on the squashing modulus without computing any functional determinant.

Our discussion in this paper was restricted to (twisted) vector supermultiplets in 3d.  It would be interesting
to extend our results to (twisted) chiral matter. To do this it would be necessary to work out the coupling of topological chiral matter to topological gravitational backgrounds: something which, to our knowledge, has not be done yet\footnote{The BRST structure of {\it rigid} topological chiral matter in 3d has been described in \cite{Ohta:2012ev}.}. The dependence on the Seifert moduli of the quantum effective action of chiral theories is considerably more complicated than that of vector supermultiplets. We expect therefore that the coupling of topological
chiral theories to topological gravitational backgrounds  involves some new ingredients. It would be equally interesting to apply our methods to higher dimensions.  Realizing this program might reduce the computation of the dependence on the moduli of quantum effective actions of localizable theories to the solution of appropriate anomalous Ward identities. We hope to come back to these problems
in the future.

\section*{Acknowledgments}
D.R. would like to thank B.~Assel, S.J.~Rey, A.~Tomasiello, F.~Yagi and A.~Zaffaroni  for interesting discussions. This work was supported in part by  INFN, by Genoa University Research Project (P.R.A.) 2014 and by the National Research Foundation of Korea grants 2005-0093843, 2010-220-C00003 and 2012K2A1A9055280.
\newpage

\appendix

 \section{Squashed 3-spheres in the Hopf coordinates}
\label{App:AppendixSquashed}
Let us start from the beginning and construct the squashed metric from $C^2$ 
 \bea
 ds^2  = |dz_1|^2  +  |dz_2|^2
 \eea
 where
 \bea
 z_1 = \rho_1\, {\rm e}^{i\,\phi_1}\qquad z_2 = \rho_2\, {\rm e}^{i\,\phi_2}
 \eea
 The $S_3$ is obtained by embedding the hypersurface
  \bea
 \rho^2 \equiv  \frac{\rho_1^2}{l^2} + \frac{\rho_2^2}{\tilde{l}^2} = 1 
 \label{hypersurface}
 \eea
 in $C^2$ , which gives
 \bea
 \rho_1 =l\, \cos\theta\qquad \rho_2 =\tilde{l}\, \sin\theta\qquad 0\le \theta\le \frac{\pi}{2}
 \eea
 Hence
 \bea
&& ds^2 = \sum_{i=1,2} |d\rho_i + i\, \rho_i\, d\phi_i|^2 = d\rho_1^2 + d\rho_2^2 + \rho_1^2\, d\phi_1^2 + \rho_2^2 \, d\phi_2^2=\nn\\
 && \qquad =\bigl( l^2\, \sin^2\theta+ \tilde{l}^2\, \cos^2\theta\bigr)\, d\theta^2 +l^2\, \cos^2\theta\, d\phi_1^2 + \tilde{l}^2\,\sin^2\theta \, d\phi_2^2
 \eea 
 and the Killing vector which gives the Seifert structure is 
 \bea
  \bar{\gamma}^\mu\partial_\mu= \frac{1}{l} \,\partial_{\phi_1} +  \frac{1}{\tilde{l}} \,\partial_{\phi_2}
 \eea

  \section{Gra\-vi\-tational CS action and the  Seifert WZ action}
  \label{App:AppendixCGSWZ}
  
The topological anomaly is, by definition, non-trivial: there is no local functional of the 3-dimensional
metric, trasforming as a 3-form,  whose BRST variation gives the anomaly. As a matter of fact, one has
\bea
  A^{(3)}_1 = s\, \Gamma_{GCS}^{(3)}(g)
\eea
where $\Gamma_{GCS}^{(3)}$ is the 3-dimensional gravitational Chern-Simons action
\bea
 \Gamma_{GCS}^{(3)}(g) = \mathrm{Tr}\, \bigl( \frac{1}{2}   \Gamma\, d\,\Gamma + \frac{1}{3}\, \Gamma^3\bigr)
 \label{GCSactionB}
\eea
Since $ \Gamma_{GCS}^{(3)}(g)$ is not a globally defined 3-form on $M_3$,  $ A^{(3)}_1$ is indeed non-trivial.

Let us discuss the relation between the Seifert Wess-Zumino action (\ref{WZSeifertaction}) and the 
gravitational CS action (\ref{GCSaction}).  Let us introduce the 3-form
\bea
\Gamma_{WZ}^{(3)\; Seif}=  \Gamma^{Seif}_{WZ}\, \, dy\, \frac{1}{2}\,\epsilon_{ij} dx^i\,dx^j
\eea
Then of course
\bea
s\, \int_{M_3}\bigl(\Gamma_{WZ}^{(3)\; Seif} - \Gamma_{GCS}^{(3)}[ g_{ij}, \sigma, a_i]\bigr)=0
\eea
where
\bea
 \Gamma_{GCS}^{(3)}[ g_{ij}, \sigma, a_i]\equiv \Gamma_{GCS}^{(3)}[g]\big|_{\mathcal{L}_{\bar{\gamma}}\, g} 
\eea
is the gravitational CS action evaluated for the metric adapted to the Seifert structure (\ref{adaptedmetric}).

Therefore, there exists a local 2-form $\Omega^{(2)}$ such that
\bea
\Gamma_{WZ}^{(3)\; Seif}- \Gamma_{GCS}^{(3)}[ g_{ij}, \sigma, a_i] = d\, \Omega^{(2)}
\eea
It turns out that 
\bea
\Omega^{(2)} = \frac{1}{4} \, \sqrt{g}\, f\, \epsilon_{ij}\, g^{jk}\, \Gamma^i_{lk}\, dx^l\, dy
\eea
$\Gamma_{GCS}^{(3)}$ is not a globally defined 3-form: This fact is expressed by
\bea
S_{\mathrm{diff}}\, \Gamma_{GCS}^{(3)} = d\, Q^{(2)}_1
\eea
where $S_{\mathrm{diff}}$ is the BRST operator associated to 3-dimensional diffeomorphisms:
\bea
S_{\mathrm{diff}} = \mathcal{L}_\xi - \{i_\xi, d\}
\eea
with $\xi= \xi^\mu\,\partial_\mu$  the reparametrization ghost in 3-dimensions.  $Q^{(2)}_1$ is the reparametrization anomaly in 2 dimensions:
\bea
Q^{(2)}_1= \frac{1}{2}\, \mathrm{Tr} \, \mathrm{M}\, d\,\Gamma= \frac{1}{2}\, \partial_\mu\,\xi^\nu d\,\Gamma^\mu_{\;\nu} 
\eea
$\Gamma_{WZ}^{(3)\; Seif}$ also is not invariant under $S_{\mathrm{diff}}$. However
we can consider reparametrization ghosts $\xi^\mu$ which are invariant under the Reeb vector $\bar{\gamma}^\mu$. Let us denote by $S^{Seif}_{\mathrm{diff}}$ the reparametrization BRST operator associated to such invariant reparametrization ghosts.  The symmetry associated to   $S^{Seif}_{\mathrm{diff}}$ is the one corresponding to
2-dimensional diffeomorphisms and abelian gauge transformations.

Although $ \Gamma_{GCS}^{(3)}[ g_{ij}, \sigma, a_i]$ is not invariant even under the restricted $\bar{S}_{\mathrm{diff}}$,  $\Gamma_{WZ}^{(3)\; Seif}$  is invariant:
\bea
S^{Seif}_{\mathrm{diff}}\,  \Gamma_{WZ}^{(3)\; Seif}=0
\eea
as it is manifest from (\ref{WZSeifertaction}). This equations expresses the fact that $\Gamma_{WZ}^{(3)\; Seif}$  is a form under reparametrizations which do not change the Seifert structure.  Hence
\bea
&& S^{Seif}_{\mathrm{diff}}\, \bigl(\Gamma_{GCS}^{(3)}[ g_{ij}, \sigma, a_i]+d\, \Omega^{(2)}\bigr)= d\,\bigl( Q^{(2)}_1+S^{Seif}_{\mathrm{diff}}\, \Omega^{(2)}\bigr)
\eea
We conclude that
\bea
 Q^{(2)}_1=S^{Seif}_{\mathrm{diff}}\, \Omega^{(2)} + d\, \Omega^{(1)}_1
 \eea
 In other words the functional $\Omega^{(2)}$ trivializes the 2-dimensional gravitational anomaly: this is possibile since $\Omega^{(2)}$ is a functional not only of the 2-dimensional metric $g_{ij}$ but also of the abelian field strength
 $f$. 

 \section{Chern-Simons  supergravity actions}
  \label{App:AppendixSUGRA}

The bosonic fields of 3d $N=2$ supergravity include, beyond the metric $g_{\mu\nu}$,   two gauge fields
$A_\mu$, $V_\mu$ and a scalar $H$.  Let us denote by $\Gamma[g_{\mu\nu}, A_\mu, V_\mu, H]$ the (non-local) effective
 action renormalized according to a prescription which preserves {\it full}  reparametrizations and gauge invariance\footnote{By ``full'' invariance we mean invariance under both ``small'' and ``large'' gauge  and coordinate transformations.}.
For matter theories which involve only vector multiplets, the dependence on the supergravity bosonic backgrounds
can be removed by subtracting from  $\Gamma[g_{\mu\nu}, A_\mu, V_\mu, H]$ a {\it local} Chern-Simons supergravity  action 
  \bea
&&  \Gamma_{c.t.}[g_{\mu\nu}, A_\mu, V_\mu, H] = \frac{1}{2} \int_{M_3}\bigl[ \mathrm{Tr} \bigl(\omega\, d\omega + \frac{2}{3}\, \omega^3\bigr)+ 4\, (A-\frac{3}{2}\, V)\, d\, (A- \frac{3}{2}\,V)+\nn\\
&&\qquad\qquad +(\mathrm{fermions})\bigr]
  \label{cssugra}
  \eea
where $\omega$ is the gravitational spin-connection.
The resulting effective action
\bea
\tilde{\Gamma}=\Gamma[g_{\mu\nu}, A_\mu, V_\mu, H]-\Gamma_{c.t.}[g_{\mu\nu}, A_\mu, V_\mu, H]
\eea
does not depend on $g_{\mu\nu}, A_\mu, V_\mu, H$ but it is not invariant under either ``large'' reparametrizations or ``large''  gauge transformations. 

The bosonic supergravity backgrounds corresponding to the supersymmetric squashed sphere are
  \bea
  && \bar{A}= -\frac{1}{2} (1- \frac{1}{b\, f}) \,d\phi_1 +  \frac{1}{2} (1- \frac{b}{f}) \,d\phi_2\qquad \bar{V}=0\nn\\
  && f^2\equiv \bigl( b^{-2}\, \sin^2\theta+ b^2\, \cos^2\theta\bigr)\nn\\
&& \bar{H} = -\frac{i}{f}
\label{sugrasquashedbkg}
\eea
 The action (\ref{cssugra}) when evalutated for these backgrounds gives
  \bea
    \Gamma_{c.t.}[\bar{g}_{\mu\nu}, \bar{A}_\mu, \bar{V}_\mu, \bar{H}] = 2 \,\pi^2\, (b-\frac{1}{b})^2
\label{gaugesugract}
    \eea
This non-trivial $b$ dependence comes entirely from the gauge part of the Chern-Simons supergravity action (\ref{cssugra}): 
the gravitational part of (\ref{cssugra}) gives a $b$-independent contribution.

Since $\tilde{\Gamma}$ is independent of the bosonic backgrounds, its value for the squashed sphere does not depend on $b$. Hence the (non-local)  reparametrization invariant and gauge invariant  effective action  $\Gamma[g_{\mu\nu}, A_\mu, V_\mu, H]$
evaluated for the  backgrounds (\ref{sugrasquashedbkg}) gives 
\bea
\Gamma[\bar{g}_{\mu\nu}, \bar{A}_\mu, \bar{V}_\mu, \bar{H}]=\Gamma_{c.t.}[\bar{g}_{\mu\nu}, \bar{A}_\mu, \bar{V}_\mu, \bar{H}]=2 \,\pi^2\, (b-\frac{1}{b})^2
\eea
up to a $b$-independent constant.
However, as we explained in Section (\ref{Section:seiferttopanomaly}), neither $\Gamma[g_{\mu\nu}, A_\mu, V_\mu, H]$ nor  $\tilde{\Gamma}$ is  the effective action renormalized with the prescription which is appropriate for quantization on Seifert manifolds.
The Seifert quantum action is obtained from $\Gamma[g_{\mu\nu}, A_\mu, V_\mu, H]$ by subtracting local counterterms
which preserve invariance under reparametrizations and gauge transformations which are compatible with the
Seifert structure and, at the same time, restore the symmetry under topological transformations of metric and
gauge fields  which do not change the Seifert moduli.
 
In Section \ref{Section:seiferttopanomaly} we showed that the appropriate local gravitational counterterm is
proportional to the Seifert Wess-Zumino action (\ref{WZSeifertaction}). In the supergravity framework we have to add to this
the analogous gauge local counterterm, to obtain\footnote{The factor 4 in front of the gauge counterterm is dictated by the presence of the same factor in (\ref{cssugra}).}
\bea
\Gamma^{Seif}_{c.t.}= \Gamma^{Seif}_{WZ}[ g_{ij}, \sigma, a_i]+ 4\,\Gamma^{Seif}_{gauge}[\phi, A^\prime]
\eea
where
\bea
\Gamma^{Seif}_{gauge}[\phi, A^\prime]= \int \bigl(\frac{1}{2} k\, dk\, \phi^2+  k \, F^\prime\, \phi\bigr)
\eea
In these formulas,  $k$ is the contact form associated to the Seifert structure
\bea
k (\bar{\gamma}) =1
\eea
$\bar{ \gamma} $ is the Reeb  Killing vector field and 
\bea
\phi= i_{\bar{\gamma}}(A) \qquad A = k\, \phi + A^\prime\qquad F^\prime = d A^\prime\qquad \mathcal{L}_{\bar{\gamma}}\,A^\prime=i_{\bar{\gamma}}(A^\prime)=0
\label{Seifertgaugebackground}
\eea
is a gauge-background compatible with the Seifert structure.

$\Gamma^{Seif}_{gauge}[\phi, A^\prime]$ is invariant under reparametrizations which preserve  the
Seifert structure.  Its topological variation reproduces the topological gauge anomaly encoded in the 3d  Abelian Chern-Simons action:
\bea
&&s\,  \Gamma^{Seif}_{gauge}[\phi, A^\prime]= \int \big[\chi\, k\,  (d\,k \,\phi+ F^\prime)+ \psi^\prime\, d\,(k\,\phi)\bigr]= \int \psi\, F\nn\\
&&s\,A = \psi= k \, \chi + \psi^\prime \qquad s\,\phi= \chi\qquad s\, A^\prime= \psi^\prime
\eea
The important point is that $\Gamma^{Seif}_{gauge}[\phi, A^\prime]$  coincides with  3d Chern-Simons action 
\bea
\Gamma_{CS}[A]= \frac{1}{2}\int A\, dA
\eea
when this is evalutated on gauge backgrounds of the Seifert form (\ref{Seifertgaugebackground})
\bea
  \Gamma_{CS}[A]= \Gamma^{Seif}_{gauge}[\phi, A^\prime]
\eea
Therefore the Seifert gauge counterterms when evaluated for the squashed sphere backgrounds gives the very same function of $b$ (\ref{gaugesugract}) which is produced by the  gauge counterterms in (\ref{cssugra})
\bea
 4\,\Gamma^{Seif}_{gauge}[\bar{\phi}, \bar{A}^\prime]=2 \,\pi^2\, (b-\frac{1}{b})^2
\eea
where $\bar{\phi}$ and $\bar{A}^\prime$ are the squashed sphere backgrounds
\bea
&&\bar{ \gamma} = b\,\partial_{\phi_1} + \frac{1}{b}\,\partial_{\phi_2}\qquad \bar{\phi} = - \frac{1}{2}\, ( b- \frac{1}{b})
\eea
This means that when one evaluates the effective action renormalized with the Seifert prescription
\bea
\tilde{\Gamma}^{Seif}=\Gamma[g_{\mu\nu}, A_\mu, V_\mu, H]-\Gamma^{Seif}_{c.t.}[g_{\mu\nu}, A_\mu, V_\mu, H]
\eea
on the squashed sphere backgrounds, the dependence on $b$  associated to the gauge counterterms  drops out
and one is left with the dependence on $b$ dictated by the gravitational  Seifert Wess-Zumino  action (\ref{WZSeifertaction})
\bea
&& \tilde{\Gamma}^{Seif}[\bar{g}_{\mu\nu}, \bar{A}_\mu, \bar{V}_\mu, \bar{H}]= \Gamma[\bar{g}_{\mu\nu}, \bar{A}_\mu, \bar{V}_\mu, \bar{H}]-\Gamma^{Seif}_{c.t.}[\bar{g}_{\mu\nu}, \bar{A}_\mu, \bar{V}_\mu, \bar{H}]= \nn\\
&& \qquad = 2 \,\pi^2\, (b-\frac{1}{b})^2-  \Gamma^{Seif}_{WZ}[ \bar{g}_{ij}, \bar{\sigma}, \bar{a}_i]-2 \,\pi^2\, (b-\frac{1}{b})^2=\nn\\
&&\qquad = -  \Gamma^{Seif}_{WZ}[ \bar{g}_{ij}, \bar{\sigma}, \bar{a}_i]= -(2\,\pi)^2\, \bigl( b^2+ \frac{1}{b^2}\bigr) 
\eea
in agreement with the result we obtained in the topological framework.

\providecommand{\href}[2]{#2}

\end{document}